\RequirePackage{snapshot}
\documentclass[a4paper,12pt]{article}

\usepackage[english]{babel}
\usepackage[T1]{fontenc}
\usepackage[utf8]{inputenc}

\usepackage[a4paper,tmargin=3truecm,bmargin=3truecm,rmargin=2.5truecm,lmargin=2.5truecm,twoside,verbose=true]{geometry}

\usepackage{fabdoc,fabmacromath}
\usepackage[thmmarks,amsmath]{ntheorem}

\theoremstyle{plain}% default
\newtheorem{thm}{Theorem}[section]
\newtheorem{lemma}[thm]{Lemma}

\newtheorem{cor}[thm]{Corollary}

\theoremstyle{plain}
\theoremseparator{.}
\theorembodyfont{\upshape}
\newtheorem{defn}{Definition}[section]
\newtheorem*{rem}{Remark}

\theoremheaderfont{\itshape\mdseries}
\theorembodyfont{\upshape}
\theoremstyle{nonumberplain}
\theoremseparator{.}
\theoremsymbol{\ensuremath{\square}}
\newtheorem{proof}{Proof}

\usepackage{stmaryrd,wasysym,upgreek,latexsym}
\usepackage{lscape}
\usepackage[vflt]{floatflt}
\usepackage{afterpage}

\usepackage{ifthen,ifpdf}

\numberwithin{equation}{section}

\allowdisplaybreaks[1]

\newcommand{\lrtimes}{\super{\ltimes}{\rtimes}}
\renewcommand{\tau}{\uptau}

\newcommand{\nc}{{non-commutative}}
\newcommand{\Nc}{{Non-commutative}}
\newcommand{\GN}{\text{GN}^{2}_{\Theta}}

\begin{document}

\title{Renormalization of the Orientable\\\Nc{} Gross-Neveu Model}
\author{Fabien Vignes-Tourneret}
\date{}
\maketitle
\vspace*{-1cm}
\begin{center}
  \textit{Laboratoire de Physique Th\'eorique\footnote{Work supported by ANR grant NT05-3-43374 ``GenoPhy''.}, B\^at.\ 210, CNRS UMR 8627\\
    Universit\'e Paris XI,  F-91405 Orsay Cedex, France\\
    e-mail: \texttt{Fabien.Vignes@th.u-psud.fr}}
\end{center}

\begin{abstract}
  We prove that the \nc{} Gross-Neveu model on the two-dimensional Moyal plane is renormalizable to all orders. Despite a remaining UV/IR mixing, renormalizability can be achieved. However, in the massive case, this forces us to introduce an additional counterterm of the form $\psib\,\imath\gamma^{0}\gamma^{1}\psi$. The massless case is renormalizable without such an addition.
\end{abstract}

%%%%%%%%%%%%%%%%%%%%%%%%%%%%%%%%
\section{Introduction}

From the rebirth of \nc{} quantum field theories
\cite{a.connes98noncom,Schomerus1999ug,Seiberg1999vs}, people were faced to
a major difficulty. A new (with respect to the usual commutative theories)
kind of divergences appeared in \nc{} field theory \cite{MiRaSe,CheRoi}. This
UV/IR mixing incited people to declare such theories non-renormalizable.
Nevertheless H.~Grosse and R.~Wulkenhhaar found recently the way to overcome
such a problem by modifying the propagator. Such a modification will be now
called ``vulcanization''. They proved the perturbative renormalizability, to
all orders, of the \nc{} $\Phi^{4}$ theory on the four-dimensional Moyal
space \cite{GrWu04-3,GrWu03-1}. Their proof is written in the matrix basis.
This is a basis for the Schwartz class functions where the Moyal product
becomes a simple matrix product \cite{Gracia-Bondia1987kw,GrWu03-2}. A Moyal
based interaction has a non-local oscillating kernel. The main advantage of
the matrix basis is that the interaction is then of the type $\Tr\Phi^{4}$.
This form is much easier to use to get useful bounds. The main drawback is the
very complicated propagator (see \cite{toolbox05} for a complete study of the
Gross-Neveu propagator in the matrix basis). This is one of the reasons which
lead us to recover in a simplified manner the
renormalizability of the \nc{} $\Phi^{4}$ theory in $x$-space \cite{xphi4-05}.
The direct space has several advantages. First of all, the propagator may be
computed exactly (and used). It has a Mehler-like form in the $\Phi^{4}$, LSZ
and Gross-Neveu theories \cite{toolbox05,xphi4-05,simon79funct}. The $x$-space
allows to compare the behaviour of commutative and \nc{} theories. It seems to
allow a simpler handling of symmetries like parity of integrals. This point is
very useful for the renormalization of the Gross-Neveu model. We also plan to
extend renormalizability proofs into the non-perturbative domain thanks to
constructive techniques developed in $x$-space. Finally, when we will be able
to do Physics with such \nc{} models, we would like to have some experience
with our physical space. Of course $x$-space has also drawbacks. It forces to
deal with non absolutely convergent integrals. We have to take care of
oscillations. Until now it is much more difficult to get the exact topological
power counting of the known \nc{} field theories in direct space than in the
matrix basis. The \nc{} parametric representation would certainly provide an
other way to get the full power counting \cite{gurauhypersyman}.\\

Apart from the $\Phi^{4}_{4}$, the modified Bosonic LSZ model \cite{xphi4-05}
and supersymmetric theories, we now know several renormalizable \nc{} field
theories. Nevertheless they either are super-renormalizable ($\Phi^{4}_{2}$
\cite{GrWu03-2}) or (and) studied at a special point in the parameter space
where they are solvable ($\Phi^{3}_{2},\Phi^{3}_{4}$
\cite{Grosse2005ig,Grosse2006qv}, the LSZ models
\cite{Langmann2003if,Langmann2003cg,Langmann2002ai}). Although only
logarithmically divergent for parity reasons, the \nc{} Gross-Neveu model is a
just renormalizable quantum field theory as $\Phi^{4}_{4}$. One of its main
interesting features is that it can be interpreted as a non-local
Fermionic field theory in a constant magnetic background. Then apart from
strengthening the ``vulcanization'' procedure to get renormalizable \nc{} field
theories, the Gross-Neveu model may also be useful for the study of the
quantum Hall effect. It is also a good first candidate for a constructive
study \cite{Riv1} of a \nc{} field theory as Fermionic models are usually
easier to construct. Moreover its commutative counterpart being asymptotically
free and exhibiting dynamical mass generation
\cite{Mitter1974cy,Gross1974jv,KMR}, a study of the physics of this model
would be interesting.\\

In this paper, we prove the renormalizability of the \nc{} Gross-Neveu model to all orders. For only technical reasons, we restrict ourselves to the orientable case. An interesting feature of the model is a kind of remaining UV/IR mixing. Some (logarithmically) divergent graphs entering the four-point function are not renormalizable by a ``local'' counterterm\footnote{By ``local'' we mean ``of the form of the initial vertex''.}. Nevertheless these ``critical'' components only appear as sub-divergences of two-point graphs. It turns out that the renormalization of the two-point function make the (four-point) critical graphs finite. In the massive case, we have to add to the Lagrangian a counterterm of the form $\delta m\,\psib\imath\gamma^{0}\gamma^{1}\psi$. The massless model is also renormalizable without such a counterterm.

In section \ref{sec:model-notations}, we present the model and fix the notations. We state our main result. Section \ref{sec:from-oscill-decr} is devoted to the main technical difficulty of the proof. Here is explained how to exploit properly the vertex oscillations in order to get the power counting. In section \ref{sec:multiscaleGN}, we compute this power counting with a multiscale analysis. In section \ref{sec:renorm-GN}, we prove that all the divergent subgraphs can be renormalized by counterterms of the form of the initial Lagrangian. Finally, appendices follow about technical details and additional properties. 

\paragraph{Acknowledgement} I am very grateful to J.~Magnen for constant discussions and critical comments. In particular he found how to use properly the vertex oscillations. I also thank V.~Rivasseau and R.~Gurau for
enlightening discussions at various stages of this work and J.-C.~Wallet for careful reading.

%%%%%%%%%%%%%%%%%%%%%%%%%%%%%%%
\section{Model and notations}
\label{sec:model-notations}

The \nc{} Gross-Neveu model ($\GN$) consists in a Fermionic quartically
interacting field theory on the (two-dimensionnal) Moyal plane
$\R^{2}_{\Theta}$. The algebra $\cA_{\Theta}$ of ``functions on
$\R^{2}_{\Theta}$'' may be defined as $\cS(\R^{2})$ (it may also be extended to an algebra of tempered distributions, see
\cite{gayral05,Gayral2003dm,Gracia-Bondia1987kw,Varilly1988jk} for rigorous
descriptions) endowed with the associative \nc{} Moyal product:
\begin{align}
  \lbt
  f\star_{\Theta}g\rbt(x)=&(2\pi)^{-2}\int_{\R^{2}}\int_{\R^{2}}dydk\,f(x+{\textstyle\frac
  12}\Theta k)g(x+y)e^{\imath k\cdot y}
\end{align}
The skew-symmetric matrix $\Theta$ is
\begin{align}
  \Theta=&
  \begin{pmatrix}
    0&-\theta\\\theta&0
  \end{pmatrix}
\end{align}
where $\theta$ is a real parameter of dimension length$^{2}$. The action
of the \nc{} Gross-Neveu model is
\begin{align}\label{eq:actfunct}
  S[\psib,\psi]=&\int
  dx\lbt\psib\lbt-\imath\slashed{\partial}+\Omega\xts+m+\imath\delta m\,\theta\gamma\Theta^{-1}\gamma\rbt\psi+V_{\text{o}}(\psib,\psi)
  +V_{\text{no}}(\psib,\psi)\rbt(x)
\end{align}
where $\xt=2\Theta^{-1}x$ and $V=V_{\text{o}}+V_{\text{no}}$ is the
interaction part given later. The term in $\delta m$ will be treated perturbatively as a counterterm. It appears from the two-loop order (see section \ref{critic-comp-2pts}). Throughout this paper we use the Euclidean
metric and the Feynman convention $\slashed{a}=\gamma^{\mu}a_{\mu}$. The
matrices $\gamma^{0}$ and $\gamma^{1}$ constitute a two-dimensionnal
representation of the Clifford algebra
$\{\gamma^{\mu},\gamma^{\nu}\}=-2\delta^{\mu\nu}$. Note that with such a
convention the $\gamma^{\mu}$'s are skew-Hermitian:
$\gamma^{\mu\dagger}=-\gamma^{\mu}$.

\paragraph{Propagator}
The propagator of the theory is given by the following lemma:
\begin{lemma}[Propagator 1 \cite{toolbox05}]\label{xpropa1}
  The propagator of the Gross-Neveu model is
  \begin{align}
    C(x,y)=&\lbt-\imath\slashed{\partial}+\Omega\xts+m\rbt^{-1}(x,y)\\
    =&\ \int_{0}^{\infty}dt\, C(t;x,y),\notag\\
    C(t;x,y)=&\ -\frac{\Omega}{\theta\pi}\frac{e^{-tm^{2}}}{\sinh(2\Ot t)}\,
    e^{-\frac{\Ot}{2}\coth(2\Ot t)(x-y)^{2}+\imath\Omega x\wed y}\\
    &\times\lb\imath\Ot\coth(2\Ot t)(\xs-\ys)+\Omega(\xts-\yts)-m\rb
    e^{-2\imath\Omega t\gamma\Theta^{-1}\gamma}\notag
  \end{align}
  with $\Ot=\frac{2\Omega}{\theta}$ and $x\wed y=2x\Theta^{-1}y$.\\
We also have $e^{-2\imath\Omega t\gamma\Theta^{-1}\gamma}=\cosh(2\Ot t)\mathds{1}_{2}-\imath\frac{\theta}{2}\sinh(2\Ot
  t)\gamma\Theta^{-1}\gamma$.
\end{lemma}
The propagator may also be considered as diagonal in some \emph{color} space indices
if we want to study $N$ copies of spin $\frac 12$ fermions.

\paragraph{Interactions}
Concerning the interaction part $V$, first remind that $\forall
f_{1},f_{2},f_{3},f_{4}\in\cA_{\Theta}$,
\begin{align}
  \int dx\,\lbt f_{1}\star f_{2}\star f_{3}\star
  f_{4}\rbt(x)=&\frac{1}{\pi^{2}\det\Theta}\int\prod_{j=1}^{4}dx_{j}f_{j}(x_{j})\,
  \delta(x_{1}-x_{2}+x_{3}-x_{4})e^{-\imath\varphi},\label{eq:interaction-phi4}\\
  \varphi=&\sum_{i<j=1}^{4}(-1)^{i+j+1}x_{i}\wed x_{j}.
\end{align}
This product is non-local and only cyclically invariant. Then, in contrast to the
commutative Gross-Neveu theory for which there is only one possible (local)
interaction, the $\GN$ model exhibits, at least, six different ones:
the \emph{orientable} interactions
\begin{subequations}\label{eq:int-orient}
  \begin{align}
    V_{\text{o}}=\phantom{+}&\frac{\lambda_{1}}{4}\sum_{a,b}\int
    dx\,\lbt\psib_{a}\star\psi_{a}\star\psib_{b}\star\psi_{b}\rbt(x)\label{eq:int-o-1}\\
    +&\frac{\lambda_{2}}{4}\sum_{a,b}\int
    dx\,\lbt\psi_{a}\star\psib_{a}\star\psi_{b}\star\psib_{b}\rbt(x)\label{eq:int-o-2}\\
    +&\frac{\lambda_{3}}{4}\sum_{a,b}\int
    dx\,\lbt\psib_{a}\star\psi_{b}\star\psib_{a}\star\psi_{b}\rbt(x),&\label{eq:int-o-3}
  \end{align}
\end{subequations}
where $\psi$'s alternate with $\psib$'s and the \emph{non-orientable} interactions
\begin{subequations}\label{eq:int-nonorient}
  \begin{align}
    V_{\text{no}}=\phantom{+}&\frac{\lambda_{4}}{4}\sum_{a,b}\int
    dx\,\lbt\psib_{a}\star\psib_{b}\star\psi_{a}\star\psi_{b}\rbt(x)\label{eq:int-no-1}&\\
    +&\frac{\lambda_{5}}{4}\sum_{a,b}\int
    dx\,\lbt\psib_{a}\star\psib_{b}\star\psi_{b}\star\psi_{a}\rbt(x)\label{eq:int-no-2}\\
    +&\frac{\lambda_{6}}{4}\sum_{a,b}\int
    dx\,\lbt\psib_{a}\star\psib_{a}\star\psi_{b}\star\psi_{b}\rbt(x).\label{eq:int-no-3}
  \end{align}
\end{subequations}
All these interactions have the same $x$-kernel thanks to
\eqref{eq:interaction-phi4}. The indices $a,b$ are spin indices taking value
in $\{0,1\}$ (or $\{\uparrow,\downarrow\}$). They may be additionnally color
indices between $1$ and $N$. For only technical reasons, we will restrict
ourselves to orientable interactions. Such a qualification will become clear
in the next section. This paper is mainly devoted to the proof of
\begin{thm}[BPHZ Theorem for $\GN$]
  The quantum field theory defined by the action
  \eqref{eq:actfunct} with $V=V_{\text{o}}$ is renormalizable to all orders of
  perturbation theory.
\end{thm}

\paragraph{Multi-scale analysis} In the following we use a multi-scale analysis \cite{Riv1}. The first step
consists in slicing the propagator as
\begin{align}
  C_{l}=&\sum_{i=0}^{\infty}C^{i}_{l},\,C^{i}_{l}=
  \begin{cases}
    {\displaystyle\int_{M^{-2i}}^{M^{-2(i-1)}}dt\,C_{l}(t;\phantom{u})}&\text{if }i\ges 1\vspace{.3cm}\\
     {\displaystyle\int_{1}^{\infty}dt\,C_{l}(t;\phantom{u})}&\text{if }i=0.\\
  \end{cases}
\end{align}
We have an associated decomposition of any amplitude of the theory as
\begin{align}
  \label{eq:scale-attrib}
  A_{G}=&\sum_{\mu}A_{G}^{\mu}
\end{align}
where $\mu=\{i_{l}\}$ runs over all possible attributions of a positive integer $i_{l}$ for each line $l$ in $G$. This index represents the ``scale'' of the line $l$. The usual ultraviolet divergences of field theory becomes, in the multi-scale framework, the divergence of the sum over attributions $\mu$ of indices. To work with well-defined quantities, we put an ultraviolet cut-off $\rho$: $i\in\{0,\dotsc,\rho\}$. In each slice, the following lemma gives a bound on the propagator.
\begin{lemma}
  For all $i\in\N$, there exists $K,\,k\in\R_{+}$ such that
  \begin{align}
    \label{eq:propabound}
    \labs C^{i}(x,y)\rabs\les&K M^{i}e^{-kM^{i}\labs x-y\rabs}.
  \end{align}
  This bound also holds in the case $m=0$.
\end{lemma}
To any assignment $\mu$ and scale $i$
are associated the standard connected components $G_k^i,\,k\in\{1,\dotsc,k(i)\}$ 
of the subgraph $G^i$ made of all lines with scales $j\ges i$. These tree components 
are partially ordered according to their inclusion relations and the (abstract) tree
describing these inclusion relations is called the Gallavotti-Nicol\`o tree \cite{GaN}; 
its nodes are the $G_k^i$'s and its root is the complete graph $G$.\\

More precisely for an arbitrary subgraph $g$ one defines: 
\begin{equation}
 i_g(\mu)=\inf_{l\in g}i_l(\mu),\quad 
 e_g(\mu)=\sup_{l \mathrm{~external~line~of~} g}i_l(\mu).
\end{equation}
The subgraph $g$ is a $G_k^i$ for a given $\mu$ if and 
only if $i_g(\mu)\ges i> e_g(\mu)$. 
As is well known in the commutative field theory case, the key to optimize the bound over 
spatial integrations is to choose the real tree $\cT$ compatible with the abstract Gallavotti-Nicol\`o tree,
which means that the restriction $\cT_k^{i}$ of  $\cT$ to any $G_k^{i}$ must still span $G_k^{i}$.
This is always possible (by a simple induction from leaves to root).\\

Let us define $i_{\nu}(\mu)$ as the index of the line of highest scale hooked to the vertex $\nu$. Then any (amputed) $N$-point function $S$ has an ``effective'' expansion:
\begin{align}
  S_{N}(x_{1},\dotsc,x_{N};\rho)=&\sum_{\text{$N$-point graphs }G}\ \sum_{\mu(G)}\prod_{\nu\in G}\lambda_{i_{\nu}}A_{G}^{\mu}(x_{1},\dotsc,x_{N};\rho).
\end{align}
Strictly speaking, we prove here that all the orders of the effective series are finite as the cut-off goes to infinity and that there exists a constant $K\in\R$ such that:
\begin{align}
  \lim_{\rho\to\infty}\int_{\R^{2N}}\prod_{i=1}^{N}dx_{i}f_{i}(x_{i})\,\labs A_{G}^{\mu}(x_{1},\dotsc,x_{N};\rho)\rabs\les K^{n(G)}
\end{align}
where the $f_{i},\,i\in\lnat 1,N\rnat$ are test functions.

\subsection{Orientation and graph variables}
\label{sec:graph-orient}

The delta function in \eqref{eq:interaction-phi4} implies that the vertex is
parallelogram shaped. To simplify the graphs, we will nevertheless draw it
either as a lozenge (Fig. \ref{fig:vertex}) or as a square.\\

\begin{floatingfigure}[p]{2cm}
  \centering
  \includegraphics[scale=1]{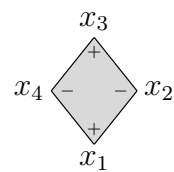}
  \caption[A vertex]{A vertex}
  \label{fig:vertex}
\end{floatingfigure}
\noindent
We associate a sign, $+$ ou $-$, to each of the four positions at a vertex.
This sign changes from a position to its neighbouring one and reflects the
signs entering the delta function. For example, the delta function associated
to the vertex of figure \ref{fig:vertex} has to be thought to be $\delta(x_{1}-x_{2}+x_{3}-x_{4})$
and not $\delta(-x_{1}+x_{2}-x_{3}+x_{4})$.  The vertex being cyclically
invariant, we can freely choose the sign of one among the four positions. The
three other signs are then fixed. Let us call \textbf{orientable} a line
joining a point $+$ to a point $-$. On the contrary if it joins two $+$ (or
$-$), we call it \textbf{clashing}. By definition, a graph is orientable if all
its lines are orientable. We will draw orientable lines with an arrow from its
$-$ to its $+$ end. The $-$ positions are then defined as outcoming a vertex and
the $+$ ones as incoming.\\

\begin{figure}[!htbp]
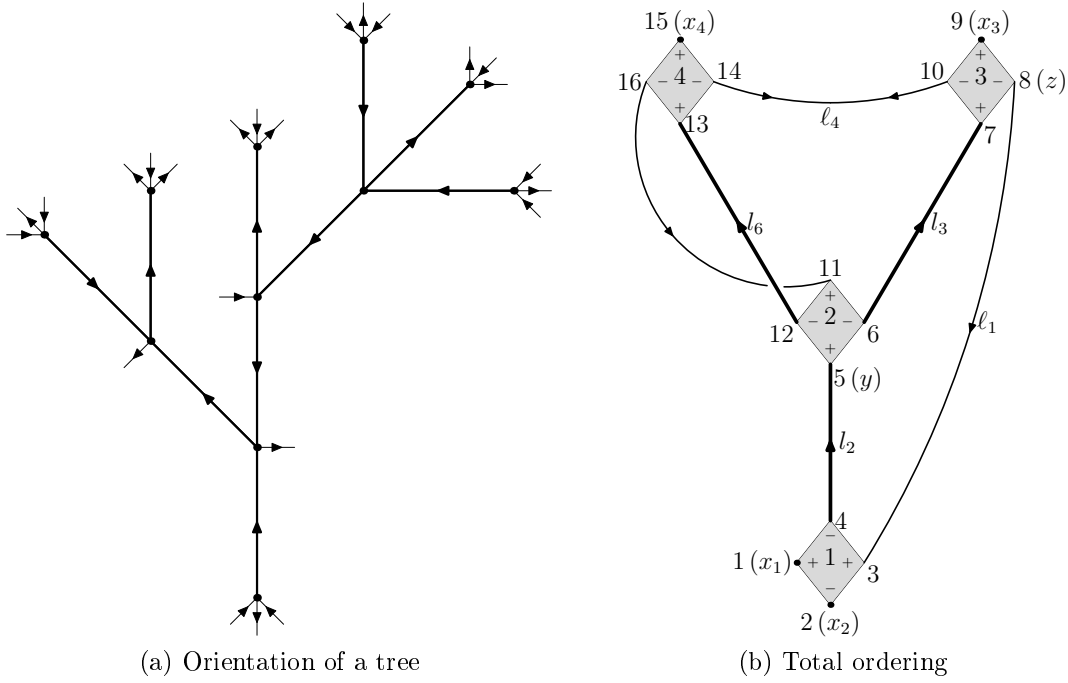

  \centering 
  \subfloat[Orientation of a tree]{{\label{fig:orientedtree}}\includegraphics[scale=1]{figures.12}}\qquad
  \subfloat[Total ordering]{\label{fig:totalorder}\includegraphics[scale=.8]{figures.45}}
  \caption{Orientability and ordering}
  \label{fig:trees}
\end{figure}

Let a graph $G$. We first choose a (optimal) spanning rooted tree $\cT$. The complete
orientation of the graph, which corresponds to the choice of the signs at each
vertex, is fixed by the orientation of the tree. For the root vertex, we
choose an arbitrary position to which we give a $+$ sign. If the graph is not
a vacuum graph, it is convenient to choose an exernal field for this reference
position. We orient then all the lines of the tree and all the remaining
half-loop lines or ``loop fields'', following the cyclicity of the vertices.
This means that starting from an arbitrary reference orientation at the root
and inductively climbing into the tree, at each vertex we follow the cyclic
order to alternate incoming and outcoming lines as in Figure
\ref{fig:orientedtree} (where the vertices are pictured as points). Let us
remark that with such a procedure, a tree is always orientable (and oriented).
The loop lines may now be orientable or not.
\begin{defn}[Sets of lines]\label{defn:sets}
  We define\\

  \ensuremath{\begin{array}{lcl}
      \cT&=&\lb\text{tree lines}\rb,\\
      \cL&=&\lb\text{loop lines}\rb=\cL_{0}\cup\cL_{+}\cup\cL_{-}
      \text{ with}\\ 
      \cL_{0}&=&\lb\text{loop lines } (+,-)\text{ or }(-,+)\rb,\\
      \cL_{+}&=&\lb\text{loop lines }(+,+)\rb,\\
      \cL_{-}&=&\lb\text{loop lines }(-,-)\rb.
    \end{array}}
\end{defn}

It is convenient to equip each graph with a total ordering among the vertex
variables. We start from the root and turn around the tree in the
trigonometrical sense. We number all the vertex positions in the order they
are met. See Figure \ref{fig:totalorder}. Then it is possible to order the
lines and external positions.
\begin{defn}[Order relations]\label{defn:relations}
  Let $i<j$ and $p<q$. For all lines
  $l=(i,j),\, l'=(p,q)\in\cT\cup\cL$, for all external position $x_{k}$, we define\\
  
  \ensuremath{\begin{array}{ccccrl}
      l&\prec&l'&&\text{if}&i<j<p<q\\
      l&\prec&k&&&i<j<k\\
      l&\subset&l'&&&p<i<j<q\\
      k&\subset&l&&&i<k<j\text{: ``$l$ contracts above $x_{k}$''}\\
      l&\ltimes&l'&&&i<p<j<q.
    \end{array}}\\
  
  \noindent
  We extend these definitions to the sets of lines defined in
  \ref{defn:sets}. For example, we write $\cL_{0}\ltimes\cL_{+}$
  instead of $\lb (\ell,\ell')\in\cL_{0}\times\cL_{+},\, \ell\ltimes
  \ell'\rb$. We also define the following set. Let $S_{1}$ and
  $S_{2}$ two sets of lines,
  \begin{align}
    S_{1}\lrtimes S_{2}=&\lb (l,l')\in S_{1}\times S_{2},\, l\ltimes l'\text{
      or }l \rtimes l'\rb.
  \end{align}
\end{defn}

For example, in Figure \ref{fig:totalorder}, $\ell_{1}\prec\ell_{4}$,
$l_{2}\subset\ell_{1}$, $l_{3}\succ x_{1}$.  Note also that with such sign
conventions, orientable lines always join an even ($-$) to an odd ($+$)
numbered position. It is now convenient to define new variables.  These are
relative to the lines of the graph whereas the variables used until now were
vertex variables. Each orientable line $l$ joins an outcoming position $x_{l-}$
to an incoming one $x_{l+}$. We define $\mathbf{u_{l}}=x_{l+}-x_{l-}$ as the
difference between the incoming and the outcoming position. For the clashing
lines, $u_{l}$ is also the difference between its two ends but the sign is
arbitrary and chosen in definition \ref{defn:longshort}. The $u_{l}$ are the
\emph{short} variables. The \emph{long} ones are defined as the sum of the two
ends of the lines. We write them $\mathbf{v_{l}}=x_{l+}+x_{l-}$ for tree lines
and $\mathbf{w_{\ell}}=x_{\ell+}+x_{\ell-}$ for the loops.
\begin{defn}[Short and long variables]\label{defn:longshort}
  Let $i<j$. For all line $l=(i,j)\in\cT\cup\cL$,
  \begin{align}
    u_{l}=&
    \begin{cases}
      (-1)^{i+1}s_{i}+(-1)^{j+1}s_{j}&\forall l\in\cT\cup\cL_{0},\\
      s_{i}-s_{j}&\forall l\in\cL_{+},\\
      s_{j}-s_{i}&\forall l\in\cL_{-}.
    \end{cases}\\
    v_{l}=&s_{i}+s_{j}\qquad\forall l\in\cT\\
    w_{l}=&s_{i}+s_{j}\qquad\forall l\in\cL.
  \end{align}
\end{defn}

Complex quantum field theory on Moyal spaces bears naturally two different
orientations. The first one is defined from the cyclic sign of the vertices.
This is the one we defined with the tree. The second one is related to the
complex feature of the theory: a field only contracts to its complex
conjugate. For the Gross-Neveu model, a line can also be oriented from its
$\psi$ end to its $\psib$ end. Then we are lead to define two different signs
for a same line.
\begin{defn}[Signs of a line]\label{defn:signe} 
  Let $i<j$. For all line $l=(i,j)\in\cT\cup\cL$,\\
  
  \ensuremath{\begin{array}{cclccll}
      \veps(l)&=&+1&&\forall l\in&\cT\cup\cL_{0}&\text{ if $i$ even}\\
      &=&+1&&&\cL_{-}\\
      &=&-1&&&\cT\cup\cL_{0}&\text{ if $i$ odd}\\
      &=&-1&&&\cL_{+}\\
      \\
      \epsilon(l)&=&+1&&\text{ if}&\psi(x_{i})\psib(x_{j})\\
      &=&-1&&\text{ if}&\psib(x_{i})\psi(x_{j}).
    \end{array}
  }
\end{defn}
\begin{cor}[Propagator 2]\label{xpropa2}
  From the definitions \ref{defn:longshort} and \ref{defn:signe}, the
  propagator corresponding to a line $l$ may be written as
  \begin{align}
    C_{l}(u_{l},v_{l})=&\ \int_{0}^{\infty}dt_{l}\, C(t_{l};u_{l},v_{l})\\
    C(t_{l};u_{l},v_{l})=&\ 
    \frac{\Omega}{\theta\pi}\frac{e^{-t_{l}m^{2}}}{\sinh(2\Ot t_{l})}\,
    e^{-\frac{\Ot}{2}\coth(2\Ot
      t_{l})u_{l}^{2}-\imath\frac{\Omega}{2}\epsilon(l)\veps(l)u_{l}\wed
      v_{l}}
    \label{xfullprop}\\ 
    &\times\lb\imath\Ot\coth(2\Ot
    t_{l})\epsilon(l)\veps(l)\us_{l}+\Omega\epsilon(l)\veps(l)\uts_{l}+m\rb
    e^{-2\imath\Omega t_{l}\gamma\Theta^{-1}\gamma}\nonumber
  \end{align}
  with $\Ot=\frac{2\Omega}{\theta}$ and where $v_{l}$ will be replaced by $w_{\ell}$
  if the propagator corresponds to a loop line.
\end{cor}

\subsection{Position routing}
\label{sec:resol-delta}
We give here a rule to solve in an optimal way the vertex delta
functions. In particular this will allow us to factorize the global
delta function (see (\ref{eq:interaction-phi4})) for each four-point
subgraph. There is no canonical way to do it but we can reject the
arbitrariness of the process into the choice of a tree. Then it is
convenient to introduce a \textbf{branch} system. To each tree line
$l$ we associate a branch $\mathbf{b(l)}$ containing the vertices
\emph{above} $l$. Let us define \emph{above}. At each vertex $\nu$,
there exists a unique tree line going down towards the root. We denote
it by $\mathbf{l_{\nu}}$. \textit{A contrario}, to each tree line $l$
corresponds a unique vertex $\nu$ such that $l_{\nu}=l$. We also define
$\mathbf{\cP_{\!\nu}}$ as the unique set of tree lines joining $\nu$ to
the root. Then the branch $b(l)$ is the set of vertices defined by
\begin{equation}
  \label{eq:branchset}
  b(l)=\lb \nu\in G|\,l\in\cP_{\!\nu}\rb.
\end{equation}
On Figure \ref{fig:totalorder}, the branch $b(l_{2})=\lb 2,3,4\rb$. We can now
replace the set of vertex delta functions by a new set associated to the
branches. Let a graph $G$ with $n$ vertices. A tree is made of $n-1$ lines
which give raise to $n-1$ branches. At each vertex $\nu$, we replace
$\delta_{\nu}(\sum_{i=1}^{4}(-1)^{i+1}x_{\nu_{i}})$ by $\delta_{l_{\nu}}(\sum_{\nu'\in
  b(l_{\nu})}\sum_{i=1}^{4}(-1)^{i+1}x_{\nu'_{i}})$. To complete this new system
of delta functions, we add to these $n-1$ first ones the ``root'' delta given
by $\delta_{G}(\sum_{\nu'\in G}\sum_{i=1}^{4}(-1)^{i+1}x_{\nu'_{i}})$. We have now
a new equivalent set of $n$ delta functions.\\

Let us precise the arguments of the branch delta functions in terms of short
and long variables. To this aim, we define the set $\mathbf{\kb(l)}$ of lines
contracting inside a given branch $b(l)$:
\begin{equation}
  \label{eq:branch-lines}
  \kb(l)=\lb l'=(x_{\nu},x_{\nu'})\in G|\nu,\nu'\in b(l)\rb.
\end{equation}
There also exists lines $l=(x_{\nu},x_{\nu'})$ with $\nu\in b(l)$ and
$\nu'\notin b(l)$. Moreover $b(l)$ may contain external positions. We denote by
$\mathbf{\cX(l)}$ the set made of the external positions in the branch $b(l)$
and of the ends (in $b(l)$) of lines joining $b(l)$ to an other branch.
From the definition \ref{defn:longshort} of short and long variables, for
fixed $\nu$, we have
\begin{equation}
  \label{eq:deltabranch}
  \sum_{\nu'\in
    b(l_{\nu})}\sum_{i=1}^{4}(-1)^{i+1}x_{\nu'_{i}}=\sum_{l\in(\cT\cup\cL_{0})\cap\kb(l_{\nu})}u_{l}
  +\sum_{\ell\in\cL_{+}\cap\kb(l_{\nu})}w_{\ell}-\sum_{\ell\in\cL_{-}\cap\kb(l_{\nu})}w_{\ell}+\sum_{e\in\cX(l_{\nu})}\eta(e)x_{e}
\end{equation}
where $\mathbf{\eta(e)}=1$ if the position $i$ is incoming and $-1$ if not. For example,
the delta function associated to the branch $b(l_{2})$ in the Figure \ref{fig:totalorder} is
\begin{equation}
  \label{eq:deltabranch2}
  \delta(y-z+x_{3}+x_{4}+u_{l_{3}}+u_{\ell_{5}}+u_{l_{6}}-w_{\ell_{4}}).
\end{equation}
In the same manner, the delta function of the complete branch is
\begin{equation}
  \label{eq:deltaroot-ex}
  \delta(x_{1}-x_{2}+x_{3}+x_{4}+u_{\ell_{1}}+u_{l_{2}}+u_{l_{3}}+u_{\ell_{5}}+u_{l_{6}}-w_{\ell_{4}}).
\end{equation}

Let us emphasize the particular case of $\delta_{G}$ 
\begin{equation}
  \label{eq:deltaroot}
  \delta_{G}\Big(\sum_{l\in\cT\cup\cL_{0}}u_{l}
  +\sum_{\ell\in\cL_{+}}w_{\ell}-\sum_{\ell\in\cL_{-}}w_{\ell}+\sum_{e\in\cE(G)}\eta(e)x_{e}\Big)
\end{equation}
where $\mathbf{\cE(G)}$ is the set of external points in $G$. Remark that for
an orientable graph $G$ ($\cL_{+}=\cL_{-}=\emptyset$), the root delta function
\eqref{eq:deltaroot} only contains the external points and the sum of all the
$u_{l}$ variables in $G$.\\

\begin{rem}
In the $\Phi^4$ model \cite{xphi4-05}, these delta functions were used to solve all the long tree variables
$v_{l},\,l\in\cT$. This is the optimal choice. Integrations over the long variables $v_{l}$ (or $w_{l}$) cost $M^{2i_{l}}$. Moreover the tree being chosen optimal, the $v_{l}$ are the most ``expensive'' long variables. From \eqref{eq:deltabranch}, we have
\begin{equation}
  \delta_{b(l)}\Big(\sum_{l'\in(\cT\cup\cL_{0})\cap\kb(l)}u_{l'}
  +\sum_{\ell\in\cL_{+}\cap\kb(l)}w_{\ell}-\sum_{\ell\in\cL_{-}\cap\kb(l)}w_{\ell}+\sum_{e\in\cX(l)}\eta(e)x_{e}\Big).
\end{equation}
There exists $e_{l}\in\cX(l)$ such that $x_{e_{l}}=\frac
12(\eta(e_{l})u_{l}+v_{l})$ (see definition \ref{defn:longshort}). This
external point is an end of the line $l$. Thus $\delta_{b(l)}$ gives
\begin{equation*}
  v_{l}=-\eta(e_{l})u_{l}-2\eta(e_{l})\Big(\sum_{l'\in(\cT\cup\cL_{0})\cap\kb(l)}u_{l'}
  +\sum_{\ell\in\cL_{+}\cap\kb(l)}w_{\ell}-\sum_{\ell\in\cL_{-}\cap\kb(l)}w_{\ell}+\sum_{e\in\cX(l)
    \setminus\{e_{l}\}}\eta(e)x_{e}\Big).
\end{equation*}
We have then used $n-1$ delta functions (one per tree line). The last one is
kept. It is the equivalent of the global momentum conservation in usual field
theories.
\end{rem}
Here we won't solve the branch delta functions. Instead we express them as oscillating integrals. In the orientable case, we have
\begin{align}
  \label{eq:delta-int0}
  \delta_{b(l)}\Big(\sum_{l'\in\kb(l)}u_{l'}+\sum_{e\in\cX(l)}\eta(e)x_{e}\Big)=&\int\frac{d^2p_{l}}{(2\pi)^2}\,e^{\imath p_{l}\cdot(\sum_{l'\in\kb(l)}u_{l'}+\sum_{e\in\cX(l)}\eta(e)x_{e})}.
\end{align}
After some manipulations on these oscillations (see section \ref{sec:masselottes}), we will get decreasing functions for the $v_{l}$'s
 and $p_{l}$'s. For each tree line $l$, we will integrate over $v_{l}$ and $p_{l}$, the final result being bounded by $\cO(1)$.

\section{From oscillations to decreasing functions}
\label{sec:from-oscill-decr}

In the preceding section, we decided to express all the vertex delta functions as oscillating integrals. Then we have $2$ independant variables per internal propagator. One is integrated over with the exponential decrease of the propagator (see \ref{xpropa2}). The other uses the propagator and vertices oscillations. Then it is useful to precise the oscillations in terms of the $u$'s and $v\,(w)$'s variables. This is done in section \ref{sec:rosfact}. We will see how to use the oscillations to get enough decreasing functions in section \ref{sec:masselottes}.

\subsection{The rosette factor}\label{sec:rosfact}

We have seen in the preceding section that the oscillations are expressed in
terms of the vertex variables whereas the propagators are naturally expressed
with short and long line variables. It is not very convenient to deal with two
equivalent sets of variables. We are then going to express the vertex
oscillations with the line variables.\\

In the following we call {\bfseries rosette factor} the set of all the vertex
oscillations plus the root delta function. We also distinguish tree lines $l$
and loop lines $\ell$\footnote{In case a line belongs to a set containing both
  tree and loop lines, we write it $l$.}. The first step to a complete
rewriting of the vertex oscillations is a ``tree reduction''. It consists in
expressing all tree variables in termes of $u$ and $v$ variables. Let a graph
$G$ of order $n$. It has $2(n-1)$ tree positions. The remaining $2n+2$ loop
and external variables are subsequently written $s_{j}$. By using the cyclic
symmetry of the vertices and the delta functions, we get (see \cite{xphi4-05}
for a proof):

\begin{lemma}[Tree reduction]\label{lemma:Filk1}
  The  rosette factor after the fisrt Filk move is \cite{Filk1996dm,xphi4-05}:
  \begin{align}
    &\delta(s_1-s_2+\dots-s_{2n+2}+\sum_{l\in \cT}u_l)\exp{\imath\varphi}\\
    &\notag\\
    \text{where }\varphi=&\sum_{i<j=0}^{2n+2}(-1)^{i+j+1}s_i\wed s_j+\frac
    12\sum_{l\in\cT}
    \veps(l)v_l\wed u_l+ \sum_{\cT\prec\cT}u_{l'}\wed u_{l}\notag\\
    &+\sum_{\lb l\in\cT,\, i\prec l\rb}u_l \wed (-1)^{i+1} s_i+\sum_{\lb
      l\in\cT,\, i\succ l\rb}(-1)^{i+1} s_i \wed u_l\notag.
  \end{align}
\end{lemma}
The next step is to express all the loop variables with the corresponding $u$
and $w$ variables. In \cite{xphi4-05}, we computed the result for \emph{planar
  regular} graph ($g=0$ and $B=1$, see appendix \ref{sec:topol-feynm-graphs} for graphologic definitions and also
\cite{GrWu04-3,Rivasseau2005bh}). Here we need the general
case\footnote{Strictly speaking, we only need, in this paper, the orientable
  case. Nevertheless the non-orientable one will follow.}. We now denote the
(true) external variables by $s_{j_{k}},\,k\in\lnat 1,N\rnat\defi \lsb
1,N\rsb\cap\N$. We write $\comp\cL_{0}\defi\cL_{+}\cup\cL_{-}$.
\begin{lemma}\label{exactoscill}
  The rosette factor of a {\bfseries general graph} is:
  \begin{align}
    &\delta\big(\sum_{k=1}^{N}(-1)^{j_{k}+1}s_{j_{k}}+\sum_{l\in\cT\cup\cL_{0}}u_l+\sum_{\ell\in\cL_{+}}w_{\ell}-\sum_{\ell\in\cL_{-}}w_{\ell}\big)%
    \,\exp\imath\varphi\\
    \nonumber\\
    \text{with }\varphi=&\ \varphi_{E}+\varphi_{X}+\varphi_{U}+\varphi_{W},\nonumber\\
    \varphi_{E}=&\ \sum_{k<l=1}^{N}(-1)^{j_{k}+j_{l}+1}s_{j_{k}}\wed s_{j_{l}},\nonumber\\
    \nonumber\\
    \varphi_{X}=&\ \sum_{k=1}^{N}\sum_{\substack{((\cT\cup\cL_{0})\prec
        j_{k})\\\cup(\comp\cL_{0}\supset j_{k})}}(-1)^{j_{k}+1}s_{j_{k}}\wed u_{l}
    +\sum_{(\cT\cup\cL_{0})\succ
      j_{k}}u_{l} \wed (-1)^{j_{k}+1}s_{j_{k}},\nonumber\\
    \nonumber\\
    \varphi_{U}=&\ \frac 12\sum_{\cT}\veps(l)v_{l}\wed u_{l}+\frac 12\sum_{\cL}\veps(\ell)w_{\ell}\wed u_{\ell}\nonumber\\
    &+\frac 12\sum_{\cL_{0}\ltimes\cL_{0}}\veps(\ell)w_{\ell}\wed
    u_{\ell'}+\veps(\ell')w_{\ell'}\wed u_{\ell}+\frac
    12\sum_{\cL_{0}\ltimes\comp\cL_0}\veps(\ell)w_{\ell}\wed
    u_{\ell'}-\veps(\ell')w_{\ell'}\wed u_{\ell}\nonumber\\
    &+\frac 12\sum_{\cL_{0}\rtimes\comp\cL_0}-\veps(\ell)w_{\ell}\wed
    u_{\ell'}+\veps(\ell')w_{\ell'}\wed u_{\ell}\nonumber\\
    &+\frac
    12\sum_{\substack{(\cL_{+}\lrtimes\cL_{-})\\\cup(\cL_{+}\ltimes\cL_{+})\cup(\cL_{-}\ltimes\cL_{-})}}u_{\ell}
    \wed\veps(\ell')w_{\ell'}+u_{\ell'}\wed\veps(\ell)w_{\ell}\nonumber\\
    &+\sum_{\substack{((\cT\cup\cL_{0})\subset\cL_{0})\\\cup((\cT\cup\cL_{0})\succ\comp\cL_0)}}\veps(\ell')
    w_{\ell'}\wed
    u_{l}+\sum_{\substack{(\comp\cL_{0}\subset\comp\cL_{0})\\
        \cup((\cT\cup\cL_{0})\prec\comp\cL_0)}}u_{l}\wed\veps(\ell')w_{\ell'}
    \nonumber\\
    &+\sum_{\substack{(\cT\cup\cL_{0})\prec(\cT\cup\cL_{0})}}u_{l'}\wed
    u_{l}+\sum_{\substack{(\cT\cup\cL_{0})\subset\comp\cL_0}}u_{l}\wed
    u_{\ell'}\nonumber\\
    &+\frac
    12\sum_{\substack{(\cL_{0}\ltimes\cL_{0})\\\cup(\cL_{+}\ltimes\cL_{+})\cup(\cL_{-}\ltimes\cL_{-})}}
    u_{\ell'}\wed u_{\ell}+\frac
    12\sum_{\substack{(\cL_{0}\lrtimes\comp\cL_0)\\\cup(\cL_{+}\rtimes\cL_{-})\cup(\cL_{-}\rtimes\cL_{+})}}u_{\ell}\wed
    u_{\ell'},\nonumber\\
    \nonumber\\
    \varphi_{W}=&\ \sum_{\substack{(\comp\cL_{0}\prec
        j_{k})\\\cup(\cL_{0}\supset j_{k})}}\veps(\ell)w_{\ell}\wed (-1)^{j_{k}+1}s_{j_{k}}
    +\sum_{\substack{\comp\cL_{0}\succ
        j_{k}}}(-1)^{j_{k}+1}s_{j_{k}}\wed \veps(\ell)w_{\ell}\nonumber\\
    &+\frac
    12\sum_{\substack{(\cL_{0}\ltimes\cL_{0})\\\cup(\comp\cL_{0}\ltimes\comp\cL_0)
        \cup(\cL_{0}\lrtimes\comp\cL_0)}}\veps
    (\ell')w_{\ell'}\wed\veps(\ell)w_{\ell}+\sum_{\substack{(\cL_{0}\supset\comp\cL_0)\\\cup(\comp\cL_{0}
        \prec\comp\cL_0)}}\veps
    (\ell')w_{\ell'}\wed\veps(\ell)w_{\ell},\nonumber
  \end{align}
  where $l$($\ell$) belongs to the set on the left-hand-side.
\end{lemma}

\begin{proof}
  As explained in section \ref{sec:resol-delta}, the root $\delta$ function is
  given by
  \begin{equation}
    \delta\big(\sum_{k=1}^{N}(-1)^{j_{k}+1}s_{j_{k}}+\sum_{l\in\cT\cup\cL_{0}}u_l+\sum_{\ell\in\cL_{+}}w_{\ell}-\sum_{\ell\in\cL_{-}}w_{\ell}\big).
  \end{equation}
  We express all the loop field variables with the $u$ and $w$ variables. Then
  the quadratic term in the external variables is
  \begin{equation}
    \sum_{k<l=1}^{N}(-1)^{j_{k}+j_{l}+1}s_{j_{k}}\wed s_{j_{l}}\, .
  \end{equation}
  
  Let an external variable $s_{j_{k}}$. The linear terms with respect to
  $s_{j_{k}}$ are
  \begin{align}
    \varphi_{j_{k}}=&\sum_{i<j_{K}}(-1)^{i+1}s_i\wed (-1)^{j_{k}}s_{j_{k}}+\sum_{i>j_{k}}(-1)^{j_{k}}s_{j_{k}}\wed (-1)^{i+1}s_i\nonumber\\
    &+\sum_{\cT\succ j_{k}}(-1)^{j_{k}}s_{j_{k}}\wed u_l +\sum_{\cT\prec
      j_{k}} u_l \wed (-1)^{j_{k}}s_{j_{k}}
  \end{align}
  where the $s_{i}$'s are all loop variables. Let a loop line $\ell=(i,j)\prec j_{k}$.\\
  Its contribution to $\varphi_{j_{k}}$ is:
  \begin{align}
    \label{eq:termlin1}
    &\lsb (-1)^{i+1}s_{i}+(-1)^{j+1}s_{j}\rsb\wed (-1)^{j_{k}}s_{j_{k}}.
  \end{align}
  The result in terms of the $u_{\ell}$ and $w_{\ell}$ variables depends on
  the orientability of the loop line. From definitions \ref{defn:longshort}
  and \ref{defn:signe}, we have
  \begin{align}
    &\lsb (-1)^{i+1}s_{i}+(-1)^{j+1}s_{j}\rsb\wed (-1)^{j_{k}}s_{j_{k}}\\
    =&\ u_{\ell}\wed (-1)^{j_{k}}s_{j_{k}}&&\text{if }\ell\in\cL_{0}\nonumber\\
    =&\ -\veps(l)w_{\ell}\wed (-1)^{j_{k}}s_{j_{k}}&&\text{if
    }\ell\in\cL_{+}\cup\cL_{-}.\nonumber
  \end{align}
  In the same way, if a loop line contracts above an external variable
  $s_{j_{k}}$, its contribution to $\varphi_{j_{k}}$ is:
  \begin{align}
    \label{eq:termlin2}
    &\lsb (-1)^{i+1}s_{i}+(-1)^{j}s_{j}\rsb\wed (-1)^{j_{k}}s_{j_{k}}\\
    =&\ -\veps(l)w_{\ell}\wed (-1)^{j_{k}}s_{j_{k}}&&\text{if
    }\ell\in\cL_{0}\nonumber\\
    =&\ u_{\ell}\wed (-1)^{j_{k}}s_{j_{k}}&&\text{if
    }\ell\in\cL_{+}\cup\cL_{-}.\nonumber
  \end{align}
  Finally the linear term for $s_{j_{k}}$ is
  \begin{align}
    \varphi_{j_{k}}=&\sum_{\substack{((\cT\cup\cL_{0})\prec
        j_{k})\\\cup(\comp\cL_{0}\supset j_{k})}}u_{l}\wed
    (-1)^{j_{k}}s_{j_{k}}+\sum_{(\cT\cup\cL_{0})\succ
      j_{k}}(-1)^{j_{k}}s_{j_{k}}\wed
    u_{l}\\
    &+\sum_{\substack{(\comp\cL_{0}\prec j_{k})\\\cup(\cL_{0}\supset
        j_{k})}}(-1)^{j_{k}}s_{j_{k}}\wed \veps(\ell)w_{\ell}
    +\sum_{\substack{\comp\cL_{0}\succ j_{k}}}\veps(\ell)w_{\ell}\wed
    (-1)^{j_{k}}s_{j_{k}}.\nonumber
  \end{align}
  
  \bigskip Let us now consider a loop line $\ell=(p,q)$. Its contribution to
  the rosette factor decomposes into a ``loop-loop'' term and a
  ``tree-loop'' term. We will detail the first one, the second one being
  obtained with the same method. The loop-loop term is:
  \begin{align}
    \varphi_{ll}=&\sum_{i<p}(-1)^{i+1}s_i\wed (-1)^p
    s_p+\sum_{\substack{p<i\\i\neq q}}(-1)^ps_p\wed
    (-1)^{i+1}s_i+(-1)^{p+q+1}s_{p}\wed s_{q}\nonumber\\
    &+\sum_{\substack{i<q\\i\neq p}}(-1)^{i+1}s_i\wed (-1)^q
    s_q+\sum_{q<i}(-1)^qs_q\wed
    (-1)^{i+1}s_i\nonumber\\
    =&\sum_{i<p}(-1)^{i+1}s_i\wed [(-1)^{p}s_p+(-1)^qs_q]+\sum_{q<i}[(-1)^ps_p+(-1)^{q}s_q]\wed (-1)^{i+1}s_i\nonumber\\
    &+\sum_{p<i<q}(-1)^{i+1}s_{i}\wed [(-1)^{p+1}s_p+(-1)^q
    s_q]+(-1)^{p+q+1}s_p\wed s_q \ .
  \end{align}
  An other loop line $\ell'=(i,j)$ has now six possibilities.  It may follow
  or precede $\ell$, contain or be contained in $\ell$, cross it by the left
  or the right. Moreover the lines $\ell$ and $\ell'$ may be orientable or
  not. I will not exhibit all these different contributions but will explain
  our method thanks to two examples.\\
  \\
  \noindent
  Let $(\ell,\ell')\in\cL_{0}^{2}$ such that $\ell'\ltimes\ell$. The line
  $\ell'$ crosses $\ell$ by the left as defined in \ref{defn:relations}. The
  corresponding term is:
  \begin{align}
    &\ (-1)^{i+1}s_i\wed [(-1)^{p}s_p+(-1)^qs_q]+(-1)^{j+1}s_{j}\wed
    [(-1)^{p+1}s_p+(-1)^qs_q]\nonumber\\
    =&\ (-1)^{i+1}s_i\wed (-u_{\ell})+(-1)^{j+1}s_{j}\wed
    (-\veps(\ell)w_{\ell})\nonumber\\
    =&\ \frac 12\lbt u_{\ell}\wed u_{\ell'}+\veps(\ell')w_{\ell'}\wed
    u_{\ell}+\veps(\ell)w_{\ell}\wed u_{\ell'}+\veps(\ell)w_{\ell}\wed
    \veps(\ell')w_{\ell'}\rbt.
  \end{align}
  In the same way, if $\ell\in\cL_{0}$, $\ell'\in\cL_{+}$ such that
  $\ell\subset\ell'$, we have:
  \begin{align}
    &\ (-1)^{i+1}s_i\wed [(-1)^{p}s_p+(-1)^qs_q]+[(-1)^ps_p+(-1)^{q}s_q]\wed
    (-1)^{j+1}s_j\notag\\
    =&\ (-1)^{i+1}s_i\wed(-u_{\ell})+(-u_{\ell})\wed
    (-1)^{j+1}s_j=u_{\ell}\wed u_{\ell'}
  \end{align}
  We do the same for the other contributions and get:
  \begin{align}
    \varphi_{ll}=&\ \frac 12\sum_{\cL}\veps(\ell)w_{\ell}\wed u_{\ell}\\
    &+\sum_{\substack{(\cL_{0}\subset\cL_{0})\\\cup(\cL_{0}\succ\comp\cL_0)}}\veps(\ell')w_{\ell'}\wed
    u_{\ell}+\sum_{\substack{(\cL_{0}\prec\comp\cL_0)\cup(\comp\cL_{0}\subset\comp\cL_{0})}}
    u_{\ell}\wed\veps(\ell')w_{\ell'}\nonumber\\
    &+\frac 12\sum_{\cL_{0}\ltimes\cL_{0}}\veps(\ell)w_{\ell}\wed
    u_{\ell'}+\veps(\ell')w_{\ell'}\wed u_{\ell}+\frac
    12\sum_{\cL_{0}\ltimes\comp\cL_0}\veps(\ell)w_{\ell}\wed
    u_{\ell'}-\veps(\ell')w_{\ell'}\wed u_{\ell}\nonumber\\
    &+\frac 12\sum_{\cL_{0}\rtimes\comp\cL_0}-\veps(\ell)w_{\ell}\wed
    u_{\ell'}+\veps(\ell')w_{\ell'}\wed u_{\ell}\nonumber\\
    &+\frac
    12\sum_{\substack{(\cL_{+}\lrtimes\cL_{-})\\\cup(\cL_{+}\ltimes\cL_{+})\cup(\cL_{-}\ltimes\cL_{-})}}u_{\ell}
    \wed\veps(\ell')w_{\ell'}+u_{\ell'}\wed\veps(\ell)w_{\ell}\nonumber\\
    &+\frac
    12\sum_{\substack{(\cL_{0}\ltimes\cL_{0})\cup(\comp\cL_{0}\ltimes\comp\cL_0)\\
        \cup(\cL_{0}\lrtimes\comp\cL_0)}}\veps
    (\ell')w_{\ell'}\wed\veps(\ell)w_{\ell}+\sum_{\substack{(\cL_{0}\supset\comp\cL_0)\\\cup(\comp\cL_{0}
        \prec\comp\cL_0)}}\veps
    (\ell')w_{\ell'}\wed\veps(\ell)w_{\ell}\nonumber\\
    &+\sum_{\substack{\cL_{0}\prec\cL_{0}}}u_{\ell'}\wed
    u_{\ell}+\sum_{\substack{\cL_{0}\subset\comp\cL_0}}u_{\ell}\wed
    u_{\ell'}\nonumber\\
    &+\frac
    12\sum_{\substack{(\cL_{0}\ltimes\cL_{0})\\\cup(\cL_{+}\ltimes\cL_{+})\cup(\cL_{-}\ltimes\cL_{-})}}
    u_{\ell'}\wed u_{\ell}+\frac
    12\sum_{\substack{(\cL_{0}\lrtimes\comp\cL_0)\\\cup(\cL_{+}\rtimes\cL_{-})\cup(\cL_{-}\rtimes\cL_{+})}}u_{\ell}\wed
    u_{\ell'}\nonumber
  \end{align}
  
  The ``tree-loop'' term is:
  \begin{align}
    \varphi_{tl}=&\sum_{\lb l'\in\cT,\, l'\prec
      p\rb}u_{l'}\wed(-1)^{p}s_{p}+\sum_{\lb l'\in\cT,\, l'\succ
      p\rb}(-1)^{p}s_p\wed
    u_{l'}\\
    &+\sum_{\lb l'\in\cT,\, l'\prec q\rb}u_{l'}\wed(-1)^{q}s_{q}+\sum_{\lb
      l'\in\cT,\, l'\succ
      q\rb}(-1)^{q}s_q\wed u_{l'}\nonumber\\
    =&\sum_{\lb l'\in\cT,\, l'\prec p\rb}u_{l'}\wed[(-1)^{p}s_p+(-1)^q
    s_q]+\sum_{\lb l'\in\cT,\, l'\succ q\rb}\lsb(-1)^{p}s_p+(-1)^q s_q\rsb\wed u_{l'}\nonumber\\
    &+\sum_{\lb l'\in\cT,\, p\prec l'\prec q\rb}u_{l'}\wed\lsb(-1)^{p+1}s_p+(-1)^{q}s_q\rsb\nonumber\\
    =&\ \sum_{\substack{\cL_{0}\succ\cT}}u_{\ell}\wed
    u_{l'}+\sum_{\substack{(\cL_{0}\prec\cT)\\\cup(\comp\cL_{0}\supset\cT)}}u_{l'}\wed
    u_{\ell}\nonumber\\
    &+\sum_{\substack{(\cL_{0}\supset\cT)\\\cup(\comp\cL_{0}\prec\cT)}}\veps(\ell)w_{\ell}\wed
    u_{l'}+\sum_{\comp\cL_{0}\succ\cT}u_{l'}\wed\veps(\ell)w_{\ell}\nonumber.
  \end{align}
\end{proof}

\begin{cor}\label{sec:oscillOrient}
  The rosette factor of an {\bfseries orientable graph} is
  \begin{align}
    &\delta\big(\sum_{k=1}^{N}(-1)^{j_{k}+1}s_{j_{k}}+\sum_{l\in\cT\cup\cL}u_l\big)%
    \,\exp\imath\varphi\\
    \nonumber\\
    \text{with }\varphi=&\ \varphi_{E}+\varphi_{X}+\varphi_{U}+\varphi_{W},\nonumber\\
    \varphi_{E}=&\ \sum_{k<l=1}^{N}(-1)^{j_{k}+j_{l}+1}s_{j_{k}}\wed s_{j_{l}},\nonumber\\
    \nonumber\\
    \varphi_{X}=&\ \sum_{k=1}^{N}\sum_{\substack{(\cT\cup\cL)\prec
        j_{k}}}(-1)^{j_{k}+1}s_{j_{k}}\wed u_{l}+\sum_{(\cT\cup\cL)\succ
      j_{k}}u_{l} \wed (-1)^{j_{k}+1}s_{j_{k}},\nonumber\\
    \nonumber\\
    \varphi_{U}=&\ \frac 12\sum_{\cT}\veps(l)v_{l}\wed u_{l}+\frac 12\sum_{\cL}\veps(\ell)w_{\ell}\wed u_{\ell}\nonumber\\
    &+\frac 12\sum_{\cL\ltimes\cL}\veps(\ell)w_{\ell}\wed
    u_{\ell'}+\veps(\ell')w_{\ell'}\wed
    u_{\ell}+\sum_{\substack{(\cT\cup\cL)\subset\cL}}\veps(\ell')w_{\ell'}\wed
    u_{l}\nonumber\\
    &+\sum_{\substack{(\cT\cup\cL)\prec(\cT\cup\cL)}}u_{l'}\wed u_{l}+\frac
    12\sum_{\substack{\cL\ltimes\cL}}u_{\ell'}\wed u_{\ell},\nonumber\\
    \nonumber\\
    \varphi_{W}=&\ \sum_{\substack{\cL\supset
        j_{k}}}(-1)^{j_{k}}s_{j_{k}}\wed \veps(\ell)w_{\ell}+\frac
    12\sum_{\substack{\cL\ltimes\cL}}\veps(\ell')w_{\ell'}\wed\veps(\ell)w_{\ell}.\nonumber
  \end{align}
\end{cor}
\begin{proof}
  It is enough to set $\cL_{+}=\cL_{-}=\emptyset$ in the general expression of
  lemma \ref{exactoscill}.
\end{proof}
\begin{cor}\label{sec:oscillRG}
  Let a {\bfseries planar regular graph} ($g=0$ and $B=1$). Its rosette factor
  is \cite{xphi4-05}
  \begin{align}
    &\delta\big(\sum_{k=1}^{N}(-1)^{k+1}x_{k}+\sum_{l\in\cT\cup\cL}u_l\big)%
    \,\exp\imath\varphi\label{eq:rosette-planreg}\\
    \nonumber\\
    \text{avec }\varphi=&\ \varphi_{E}+\varphi_{X}+\varphi_{U},\nonumber\\
    \varphi_{E}=&\ \sum_{i<j=1}^{N}(-1)^{i+j+1}x_{i}\wed
    x_{j},\nonumber\\
    \nonumber\\
    \varphi_{X}=&\ \sum_{k=1}^{N}\sum_{\substack{(\cT\cup\cL)\prec
        k}}(-1)^{k+1}x_{k}\wed u_{l}+\sum_{(\cT\cup\cL)\succ k}u_{l}\wed (-1)^{k+1}x_{k},\nonumber\\
    \nonumber\\
    \varphi_{U}=&\ \frac 12\sum_{\cT}\veps(l)v_{l}\wed u_{l}+\frac
    12\sum_{\cL}\veps(\ell)w_{\ell}\wed u_{\ell}\nonumber\\
    &+\sum_{(\cT\cup\cL)\subset\cL}\veps(\ell')w_{\ell'}\wed
    u_{l}+\sum_{\substack{(\cT\cup\cL)\prec(\cT\cup\cL)}}u_{l'}\wed
    u_{l}.\nonumber
  \end{align}
\end{cor}

\begin{proof}
  As the graph has only one broken face, there is always an even number of
  fields between two external variables. In this case, $j_{k}$ and $k$ have
  the same parity. Thus by switching $s_{j_{k}}$ into $  x_{k}$, the quadratic
  term in the external variables is:
  \begin{equation}
    \sum_{i<j=1}^{N}(-1)^{i+j+1}x_{i}\wed x_{j}.
  \end{equation}
  Moreover the constraints $g=0$ and $B=1$ imply that the graph is
  orientable ($\cL=\cL_{0}$). Indeed, let us consider a clashing loop line $\ell$
  joining $s_{i}$ to $s_{i+2p}$. These two positions have same parity. Between
  the two ends of $\ell$ are an odd number of positions. Then either
  $\ell$ contracts above an external variable and $B\ges 2$, or an other loop
  line crosses it and $g\ges 1$.\\
  Finally by skipping from the result of lemma \ref{exactoscill} the terms
  concerning crossing lines, lines contracting above external variables
   and non-orientable lines, we get (\ref{eq:rosette-planreg}).
\end{proof}

\subsection{The masslets}
\label{sec:masselottes}

Contrary to the $\Phi^{4}$ case, the Gross-Neveu propagator $C^{i}$
(\ref{xfullprop}) does not contain any term of the form $\exp-M^{-2i}w^{2}$
(we call them \emph{masslets}) \cite{xphi4-05}. This term is replaced by an
oscillation of the type $u\wed w$. Whereas masslets are not in the propagator,
they appear after integration over the $u$ variables:
\begin{equation}
  \label{eq:masscreation}
  \int d^{2}u\, e^{-M^{2i}u^{2}+\imath u\wed w}=K M^{-2i}\, e^{-M^{-2i}w^{2}}.
\end{equation}
Let $G$ a connected graph. Its amplitude is
\begin{align}
  A_{G}=\int&\prod_{i=1}^{N}dx_{i}\,f_{i}(x_{i})\delta_{G}\prod_{l\in\cT}du_{l}dv_{l}\,\delta_{b(l)}C_{l}(u_{l},v_{l})
  \prod_{\ell\in\cL}du_{\ell}dw_{\ell}\,C_{\ell}(u_{\ell},w_{\ell})e^{i\varphi}.
\end{align}
The points $x_{i},\,i\in\lnat 1,N\rnat$ are the external positions.
For the delta functions, we use the notations of section
\ref{sec:resol-delta}. The total vertex oscillation $\varphi$  is given by the
lemma \ref{exactoscill}. It is convenient to split the propagator into two
parts. We define, for all line $l\in G$, $\mathbf{\bar{C}}_{l}(u_{l})$ by
$C_{l}(u_{l},v_{l})=\bar{C}_{l}(u_{l})\,
e^{-\imath\frac{\Omega}{2}\epsilon(l)\veps(l)u_{l}\wed v_{l}}$. Once more we
replace $v$ by $w$ for loop lines. This splitting allows to gather the
propagators oscillations with the vertex ones. The total oscillation
$\varphi_{\Omega}$ is simply deduced from $\varphi$ by replacing the terms
$\frac 12\veps(l)v_{l}\wed u_{l}$ by
$\frac 12(1+\epsilon(l)\Omega)\veps(l)v_{l}\wed u_{l}$. The graph amplitude
becomes
\begin{align}
  A_{G}=\int&\prod_{i=1}^{N}dx_{i}\,f_{i}(x_{i})\delta_{G}\prod_{l\in\cT}du_{l}dv_{l}\,\delta_{b(l)}\bar{C}_{l}(u_{l})
  \prod_{\ell\in\cL}du_{\ell}dw_{\ell}\,\bar{C}_{\ell}(u_{\ell})e^{i\varphi_{\Omega}}.
\end{align}
In contrast with the $\Phi^4$ theory \cite{xphi4-05}, we won't solve the branch delta functions. Instead we keep $\delta_{G}$ but express the $n-1$ other delta functions as oscillating integrals:
\begin{align}
  \label{eq:delta-int}
  \delta_{b(l)}\Big(\sum_{l'\in\kb(l)}u_{l'}+\sum_{e\in\cX(l)}\eta(e)x_{e}\Big)=&\int\frac{d^2p_{l}}{(2\pi)^2}\,e^{\imath p_{l}\cdot(\sum_{l'\in\kb(l)}u_{l}+\sum_{e\in\cX(l)}\eta(e)x_{e})}.
\end{align}
As already explained in section \ref{sec:resol-delta}, there exists $e_{l}\in\cX(l)$ such that $x_{e_{l}}=\frac
12(\eta(e_{l})u_{l}+v_{l})$. Remark that $\eta(e_{l})=\veps(l)$. Then
\begin{align}
  \sum_{l'\in\kb(l)}u_{l'}+\sum_{e\in\cX(l)}\eta(e)x_{e}=&\frac 12(u_{l}+\veps(l)v_{l})+\sum_{l'\in\kb(l)}u_{l'}+\sum_{e\in\cX(l)\setminus\{e_{l}\}}\eta(e)x_{e}.
\end{align}
In the following we will use an additionnal notation. For all line $l\in\cT$, let us define $\nu_{l}$ as the unique vertex such that $l=l_{\nu}$ where $l_{\nu}$ is defined in section \ref{sec:resol-delta}. $\nu_{l}$ is the vertex just above $l$ in the tree. We write $\varphi'_{\Omega}$ for the total oscillation where we add the new oscillations resulting from the delta functions\footnote{Note that the oscillation is invariant under $p_{l}\to -p_{l}$ for all $l\in G$ independently.}. The graph
amplitude is now
\begin{align}
  A_{G}=\int&\prod_{i=1}^{N}dx_{i}\,f_{i}(x_{i})\delta_{G}\prod_{l\in\cT}du_{l}dv_{l}dp_{l}\,\bar{C}_{l}(u_{l})
  \prod_{\ell\in\cL}du_{\ell}dw_{\ell}\,\bar{C}_{\ell}(u_{\ell})e^{i\varphi'_{\Omega}}.\label{eq:amplitude-avt-massel}
\end{align}
Remark that we have omitted the factors $2\pi$ as we have done until now and will go
on doing with the $\frac{-\lambda}{4\pi^{2}\det\Theta}$ vertex factors.
To get the masslets, we could, for example, integrate over the variables
$u_{l}$.  This exact computation would be the equivalent of
equation \eqref{eq:masscreation}. We should integrate $2n-N/2$ coupled
Gaussian functions. We would get Gaussian functions in some variables
$\cW_{l}$ which would be linear combinations of
$w_{\ell'}$.  Apart from the difficulty of this computation, we should
then prove that the obtained decreasing functions are independant. For
general graphs, it is somewhat difficult. Then instead of computing an
exact result, we get round the difficulty by exploiting the
oscillations before integrating over the $u$'s, $v$'s and $w$'s. The rest of
this section is devoted to the proof of
\begin{lemma}\label{lem:masselottes}
  Let $G$ an orientable graph with $n$ vertices and $\mu$ a scale attribution. For all $\Omega\in\lsb 0,1\rbt$, there exists
  $K\in\R$ such that the amplitude (\ref{eq:amplitude-avt-massel}), amputed,
  integrated over test functions, with the $\mu$ attribution, is bounded uniformly in $n$ by
  \begin{align}
    \labs A^{\mu}_{G}\rabs\les&K^{n}\int
    dx_{1}\,g_{1}(x_{1}+\{a\})\delta_{G}\prod_{i=2}^{N}dx_{i}\,g_{i}(x_{i})\prod_{l\in G}da_{l}\,M^{2i_{l}}\Xi(a_{l})\\
    &\qquad\prod_{l\in\cT}du_{l}d\cV_{l}dp_{l}\,M^{i_{l}}e^{-M^{2i_{l}}(u_{l}-\veps(l)a_{l})^{2}}\prod_{\mu=0}^{1}\frac{1}{1+M^{-2i_{l}}\cV^{2}_{l,\mu}}
    \frac{1}{1+M^{2i_{l}}p^{2}_{l,\mu}}\notag\\
    &\qquad\prod_{\ell\in\cL}
    du_{\ell}d\cW_{\ell}M^{i_{\ell}}
    e^{-M^{2i_{\ell}}(u_{\ell}+\{a\})^{2}}
    \prod_{\mu=0}^{1}\frac{1}{1+M^{-2i_{\ell}}\cW^{2}_{\ell,\mu}}\notag\\
    \text{with }\veps(l)\cV_{l}=&{\textstyle\frac
    12}(1+\epsilon(l)\Omega)\veps(l)v_{l}+\sum_{\ell'\supset l}\veps(\ell')w_{\ell'}-{\textstyle\frac 12}\pt_{l}-\sum_{l'\in\cP_{\kv_{l}}}\pt_{l'},\label{eq:Vl}\\
    \veps(\ell)\cW_{\ell}=&{\textstyle\frac
    12}(1+\epsilon(\ell)\Omega)\veps(\ell)w_{\ell}+\sum_{\ell'\supset\ell}\veps(\ell')w_{\ell'}+
    \sum_{\ell'\ltimes\ell}\veps(\ell')w_{\ell'}\label{eq:Wl}
  \end{align}
  and $\pt=\frac 12\Theta p$, $g_{i},\,i\in\lnat 1,N\rnat$ and $\Xi$ are test functions such that $\|g_{i}\|\les\sup_{0\les p\les 2}\|f^{(p)}_{i}\|$.
\end{lemma}

\medskip Remind that we restrict our analysis to orientable graphs. We introduce a Schwartz class function $\xi\in\cS(\R^{2})$ which, conveniently scaled, is going to mimic the decrease of propagators on a scale $M^{-i_{l}}$. We want to
get a decreasing function in $v_{l}$ without integrating over
$u_{l}$. We use
\begin{align}
  1=&\int d^{2}a_{l}\, \coth(2\Ot t_{l})\xi(a_l\coth^{1/2}(2\Ot t_{l})).\label{eq:un}
\end{align}
The coupling between this $1$ and the rest of the graph is made by an \textit{ad hoc}
change of variables. We have two constraints on such a change. On one hand we
want independant decreasing functions. On the other hand, for all line $l$,
the decresase should be of scale\footnote{In some cases, a line may have a
  masslet of a scale greater than its own index. These cases are restricted to
  a single class of graphs we will detail in section \ref{subsec:brokenfaces}.} $M^{i_{l}}\simeq\coth^{1/2}(2\Ot t_{l})$.\\

We are going to make masslets line by line.  Let us write $x_{1}$ for
the root position. Let a tree line $l$. We perform the change of
variables
\begin{align}
  \lb
  \begin{aligned}
    u_{l}\to&u_{l}-\veps(l)a_{l},\\
    x_{1}\to&x_{1}+\eta(1)\veps(l)a_{l}.
  \end{aligned}\right.\label{eq:chgtvarVl}
\end{align}
It is not difficult to check that
$\varphi'_{\Omega}\to\varphi'_{\Omega}+a_{l}\wed\cV_{l}+a_{l}\wed(U_{l}+A_{l}+X_{l})$
where $\cV_{l}$ is given by (\ref{eq:Vl}) and $U_{l}$, $A_{l}$ and
$X_{l}$ are respectively linear combinations of $u$'s, $a$'s and
external variables $x$'s. Please note that such a change of variables let the global root delta function unchanged. 
Writing only the terms in the amplitude $A_{G}$ depending on $a_{l}$, we get
\begin{align}
  A_{G,l}=\int&
  da_{l}\int_{M^{-2i_{l}}}^{M^{-2(i_{l}-1)}}dt_{l}\,\coth(2\Ot
  t_{l})\xi(a_{l}\coth^{1/2}(2\Ot t_{l}))\\
  &\lb\imath\Ot\coth(2\Ot
  t_{l})(\epsilon\veps)(l)(\us_{l}-\veps(l)\slashed{a}_{l})+\Omega(\epsilon\veps)(l)
  (\uts_{l}-\epsilon(l)\slashed{\tilde{a}}_{l})-m\rb\notag\\
  &e^{-\frac{\Ot}{2}\coth(2\Ot
    t_{l})(u_{l}-\veps(l)a_{l})^{2}}f_{1}(x_{1}+\eta(1)\veps(l)a_{l})\,
 	e^{\imath a_{l}\wed(\cV_{l}+U_{l}+A_{l}+X_{l})}\notag\\
  =\int&
  da_{l}dt_{l}\,\coth(2\Ot
  t_{l})\xi(a_{l}\coth^{1/2}(2\Ot t_{l}))\,
  e^{-\frac{\Ot}{2}\coth(2\Ot
    t_{l})(u_{l}-\veps(l)a_{l})^{2}}f_{1}(x_{1}+\eta(1)\veps(l)a_{l})\notag\\
  &\hspace{-.3cm}\lb\imath\Ot\coth(2\Ot
  t_{l})(\epsilon\veps)(l)(\us_{l}-\veps(l)\slashed{a}_{l})+\Omega(\epsilon\veps)(l)
  (\uts_{l}-\veps(l)\slashed{\tilde{a}}_{l})-m\rb e^{\imath a_{l}\wed(U_{l}+A_{l}+X_{l})}\nonumber\\
  &\prod_{\mu=0}^{1}\lbt\frac{\coth^{1/2}(2\Ot t_{l})+\frac{\partial}{\partial
      a_{l}^{\mu}}}{\coth^{1/2}(2\Ot
    t_{l})+\imath\cVt_{l,\mu}}\rbt^{\!\!\!2}  e^{\imath a_{l}\wed\cV_{l}}.\label{eq:int-part}
  \intertext{We now integrate by parts over $a_{l}$. The boundary terms
  vanish. We give here the order of magnitude of the result. The details of
  the computation are given in appendix \ref{sec:integration-parts}.}
  A_{G,l}\simeq\int&da_{l}dt_{l}\,\coth(2\Ot t_{l})e^{\imath
    a_{l}\wed\cV_{l}}\prod_{\mu=0}^{1}\lbt\frac{1}{\coth^{1/2}(2\Ot
    t_{l})+\imath\cVt_{l,\mu}}\rbt^{\!\!\!2}\Xi(a_{l}\coth^{1/2}(2\Ot t_{l}))\notag\\
  &e^{-\frac{\Ot}{2}\coth(2\Ot
    t_{l})(u_{l}-\veps(l)a_{l})^{2}}e^{\imath a_{l}\wed(U_{l}+A_{l}+X_{l})} g_{1}(x_{1}+\eta(1)\veps(l)a_{l})\,\cO\big(\coth^{3/2}(2\Ot
  t)\big).\label{eq:intpart-result}
  \intertext{Then we get the following bound}
  \labs A_{G,l}\rabs\les&
  KM^{-i_{l}}e^{-kM^{2i_{l}}(u_{l}-\veps(l)a_{l})^{2}}g_{1}(x_{1}+\eta(1)\veps(l)a_{l})
  \prod_{\mu}\frac{1}{1+M^{-2i_{l}}\cV^{2}_{l,\mu}}.\label{eq:borne-interm}
\end{align}

Let us now explain how to get the corresponding decreasing functions for the $p_{l}$ variables. We begin by performing the change of variables $v_{l}\to\cV_{l}$ for all tree line $l$. The determinant of the corresponding Jacobian matrix is $2^{-(n-1)}\prod_{l\in\cT}(1+\epsilon(l)\Omega)$. It is non-vanishing for all $\Omega\in\lsb 0,1\rbt$. The total oscillation becomes
\begin{align}
  \label{eq:oscillapresVl}
  \varphi'_{\Omega}=&\varphi_{E}+\varphi_{X}+\varphi_{W}+\sum_{\cT}\veps(l)\cV_{l}\wed (u_{l}-\veps(l)a_{l})+\sum_{\cT}p_{l}\!\cdot\!(\sum_{l'\in\cL\cap\kb(l)}u_{l}+\sum_{e\in\cX(l)\setminus\{e_{l}\}}\eta(e)x_{e})\nonumber\\
    &+\sum_{\cT}(1+\epsilon(l)\Omega)^{-1}\veps(l)\cV_{l}\!\cdot\!p_{l}+WR_{1}P+PR_{2}P\notag\\
    &\hspace{-.4cm}+\frac 12\sum_{\cL}(1+\epsilon(\ell)\Omega)\veps(\ell)w_{\ell}\wed u_{\ell}+\frac 12\sum_{\cL\ltimes\cL}\veps(\ell)w_{\ell}\wed
    u_{\ell'}+\veps(\ell')w_{\ell'}\wed
    u_{\ell}+\sum_{\substack{\cL\subset\cL}}\veps(\ell')w_{\ell'}\wed
    u_{\ell}\nonumber\\
    &+\sum_{\substack{(\cT\cup\cL)\prec(\cT\cup\cL)}}u_{l'}\wed u_{l}+\frac
    12\sum_{\substack{\cL\ltimes\cL}}u_{\ell'}\wed u_{\ell}+AR_{3}A+AR_{4}U+AR_{5}X
\end{align}
where we used the notations of corollary \ref{sec:oscillOrient} and $R_{i},\,i\in\lnat 1,5\rnat$ are skew-symmetric matrices. By using
\begin{align}
  e^{\imath(1+\epsilon(l)\Omega)^{-1}\veps(l)\cV_{l}\cdot p_{l}}=&\frac{M^{-i_{l}}+(1+\epsilon(l)\Omega)\veps(l)\frac{\partial}{\partial\cV_{l,\mu}}}{M^{-i_{l}}+\imath p_{l,\mu}}e^{\imath(1+\epsilon(l)\Omega)^{-1}\veps(l)\cV_{l}\cdot p_{l}}
\end{align}
and integrating by parts over $\cV_{l}$, we get a decreasing function in $p_{l}$ which behaves like $\lbt 1+M^{2i_{l}}p_{l}^{2}\rbt^{-1}$.
We now turn to the loop lines. We also want to get decreasing functions for them. Let a loop line $\ell=(x_{\ell},x'_{\ell})\in\cL$ of $G$ with
$x_{\ell}\prec x'_{\ell}$. We make the following change of variables\footnote{This change of variables is slightly different from the one we used for the tree lines (\ref{eq:chgtvarVl}). This leads to an easier proof of the independance of the decreasing functions.}
\begin{align}
  \lb
  \begin{aligned}
    u_{\ell}\to&u_{\ell}-\veps(\ell)a_{\ell},\\
    w_{\ell}\to&w_{\ell}+a_{\ell},\\
    x_{1}\to&x_{1}+\eta(1)\veps(\ell)a_{\ell}.
  \end{aligned}\right.\label{eq:chgtvarWl}
\end{align}
The changes concerning $u_{\ell}$ and $w_{\ell}$ correspond to ``move''
$x_{\ell}$. It is easy to check that \eqref{eq:chgtvarWl} implies
$\varphi'_{\Omega}\to\varphi'_{\Omega}+a_{\ell}\wed\cW_{\ell}+a_{\ell}\wed(U_{\ell}+A_{\ell}+X_{\ell}+P_{\ell})$ where
$\cW_{\ell}$ is given by \eqref{eq:Wl} and $U_{\ell}$, $A_{\ell}$, $X_{\ell}$ and $P_{\ell}$ are respectively linear combinations of $u$'s, $a$'s, external variables $x$'s and $p$'s. We can perform the same type of integration by parts than we used for the tree variables $v_{l}$ and obtain bounds similar to (\ref{eq:borne-interm}). This proves lemma \ref{lem:masselottes}.

\paragraph{Independance of the decreasing functions}
Remind that the above procedure had two main goals. First of all we wanted to get decreasing functions of scale $i_{l}$ for all variables $v_{l}\,(w_{l})$. This should be clear from the preceding section. The second aim was the independance of those decreasing functions. Our procedure is designed to make transparent such an independance.\\

In section \ref{sec:model-notations}, definition \ref{defn:relations} gave a way to partially order the lines. This ordering was useful to express the vertex oscillations in terms of the $u$'s, $v$'s and $w$'s. But we can also define a total ordering among the lines of a graph. We say that $l<l'$ if the first end (in the trigonometric sense around the tree) of $l$ is met before the first end of $l'$. Then for all line $l\in G$, $\cV_{l}\,(\cW_{l})$ depends only on $v_{l'}$'s and $w_{l'}$'s with $l'<l$. Let $V\,(W)$ and $V'\,(W')$ the vectors containing respectively the 
variables $\veps(l)v_{l}\,(\veps(\ell)w_{\ell})$ and $\veps(l)\cV_{l}\,(\veps(\ell)\cW_{\ell})$. Let $M^{-1}$ the Jacobian matrix of the change of variables $(\veps v\ \veps w)\to(\veps\cV\ \veps\cW)$: $(V'\ W')=M(V\ W)$. The ordering introduced just above allows to prove that $M$
is triangular. Its determinant is
\begin{align}
  \det M=2^{-(2n-N/2)}\prod_{l\in G}(1+\epsilon(l)\Omega).
\end{align}
Clearly $\forall\Omega\in\lsb 0,1\rbt,\,\det M\neq 0$ and $M$ is
invertible. The decreasing functions in $\cV_{l}\,(\cW_{\ell})$ are consequently independant.
\begin{rem}
  With the non-orientable interactions (\ref{eq:int-nonorient}), we were not able to find a procedure making the independance of the masslets transparent.
\end{rem}

%%%%%%%%%%%%%%
\subsection{Non-planarity}
\label{subsec:nonplanar}

In the preceeding section, we proved that the vertex and propagators
oscillations of the Gross-Neveu model allow to obtain decreasing
functions similar to the masslets of the (\nc{}) $\Phi^{4}$
theory. Here we improve these decreases if the graph is
non-planar. For this the lemma \ref{lem:masselottes} is not
sufficient. Before taking the module of the graph
amplitude, we would like to further exploit the oscillations.\\

Let $T^{-1}$ the Jacobian matrix of the change of variables $\veps w\to\veps\cW$: $W'=TW$.
Let us define the skew-symmetric matrix $Q_{W}$ with
$\varphi_{W}=WQ_{W}W$ where $\varphi_{W}$ is given by corollary
\ref{sec:oscillOrient}. After the change of variables $W\to W'=TW$,
$\varphi_{W}=W'Q'_{W}W'$ with\\
$Q'_{W}=^{\phantom{T}t}\!\!T^{-1}Q_{W}T^{-1}$. $T$ being invertible,
the rank of $Q'_{W}$ equals $Q_{W}$'s. Remark that $Q_{W}$ is the
intersection matrix of the graph. We have the following result $\rk
Q_{W}=2g$ \cite{gurauhypersyman,CheRoi}. Let us consider a non-planar
graph.  The rank of $Q_{W}$ being different from zero, there exists a
loop line $\ell$ such that we have an oscillation
$\cW_{\ell}\wed\cW'_{\ell}$ with
$\cW'_{\ell}=\sum_{\ell'}Q'_{W,\ell\ell'}\cW_{\ell}+U+A+X+P$. Thanks to
lemma \ref{lem:masselottes}, we know that $\cW_{\ell}$ decreases on a
scale $M^{i_{\ell}}$ with the function
$(1+M^{-2i_{\ell}}\cW_{\ell}^{2})^{-1}$. By an integration by parts
similar to \eqref{eq:int-part}, we get a decrease in $\cW'_{\ell}$ on
a scale $M^{-i_{\ell}}$. This decrease will be used to integrate over
some $\cW_{\ell'}$ contained in $\cW'_{\ell}$. The result of such an
integration will be of order $M^{-2i_{\ell}}$ instead of
$M^{2i_{\ell'}}$. The gain is then $M^{-2i_{\ell}-2i_{\ell'}}$.

\subsection{Broken faces}
\label{subsec:brokenfaces}

We remind that a broken face is a face to which belongs external points (see appendix \ref{sec:topol-feynm-graphs} for examples). When
we do not consider vacuum graphs, there is always at least one broken face. By
definition, it is called the external face. The broken faces produce
oscillations of the type $x\wed w$ (see lemma \ref{exactoscill}). In the
planar case with $B\ges 2$ broken faces, we are going to use such oscillations
to get better decreases than the ones of the lemma \ref{lem:masselottes}. Let $Q_{X\!W}$ the skew-symmetric matrix representing the oscillations between the $x$'s and $w$'s variables. After the change of variables $W\to W'$, this matrix becomes
\begin{equation}
Q'_{X\!W}=Q_{X\!W}T^{-1}.
\end{equation}
Then $\rk Q'_{X\!W}=\rk Q_{X\!W}$. Let $I$ a set of consecutive natural numbers indexing some external variables
$x_{k},\, k\in I$. These ones oscillate with the variables $w_{\ell},\,\ell\in
B_{I}$ where $B_{I}$ is the set of lines contracting above those variables.
Let us now check that the variables $x_{k},\,k\in I$ oscillate only with
$\cW_{\ell},\,\ell\in B_{I}$. To this aim, let us assume that two sets $X$ and
$Y$ of external variables oscillate with two other different sets $A$ and $B$
of loop lines:
\begin{align}
  Q_{X\!W}&=
  \begin{pmatrix}
    A&0\\
    0&B
  \end{pmatrix},\quad T=
  \begin{pmatrix}
    C&0\\
    0&D
  \end{pmatrix}
\\
  Q'_{X\!W}&=Q_{X\!W}T^{-1}=
  \begin{pmatrix}
    AC^{-1}&0\\
    0&BD^{-1}
  \end{pmatrix}.
\end{align}
In the planar case, $\cW_{\ell}$ is only function of $w_{\ell'}$ with
$\ell'\supset\ell$. $T$ (and $T^{-1}$) are then not only (lower) triangular
but also bloc diagonal. The oscillations between the external variables
$x_{k}$ and the variables $\cW_{\ell}$ are
\begin{equation}
  X_{I}Q_{X\!W}T^{-1}W'_{B_{I}}=\sum_{k\in
  I}\eta(k)x_{k}\wed\text{CL}(\cW_{\ell},\,\ell\in B_{I})
\end{equation}
where $\text{CL}$ means ``linear combination''. After the masslets and non-planar
cases, it should be clear that this new oscillation allows to get a decreasing
function of scale $M^{-\min_{\ell\in B_{I}}i_{\ell}}$ in the external
variables. If these points are ``true'' external ones (of scale $-1$, integrated with test functions), we will use it to improve the power counting. Usually external points are integrated over test functions (the
result is of order $1$) so that the gain is here $M^{-2\min i_{\ell}}$.

%%%%%%%%%%%%%%%%%%%%%%%%%%%%%%%%%%%%%%%%%%%
\section{Power counting}
\label{sec:multiscaleGN}

In this section, we use the previous decreases by adapting them to the
multi-scale case.  By lemma \ref{lem:masselottes}, we know that it is
possible to get $\labs\cL\rabs$ independant decreasing functions
equivalent to the masslets of the $\Phi^{4}$ theory plus $n-1$
masslets for the tree lines coupled to $n-1$ strong decreases. These
last two types of decreasing functions are equivalent to the branch
delta functions. The method we use to get the power counting now depends on the topology of the considered graph\footnote{The main result is lemma \ref{lem:compt-puiss}, in particular in regard to the power counting of the \emph{critical} function $N=4$, $B=2$ which manages the main technical point in providing renormalizabiblity.}.\\

We only consider graphs with at least two external legs. The vacuum graphs are considered in appendix \ref{sec:vacuum-graph}.
We use the Gallavotti-Nicol\`o tree. We start from its leaves and go
down towards the root which means from the scale of the ultraviolet
cut-off to the scale $0$. Let $G^{i}_{k}$ an orientable connected
component. For all lines, we first get all the masslets by the method expounded in section \ref{sec:masselottes}. If $G^{i}_{k}$ is planar regular ($g=0$, $B=1$), we directly use lemma
\ref{lem:masselottes}. If $G^{i}_{k}$ is non-planar ($g\ges 1$), we
use the $\cW\wed\cW$ oscillations. Thanks to the procedure explained
in section \ref{subsec:nonplanar}, we get an additionnal decrease in
some $\cW'_{\ell},\,\ell\in\cL^{i}_{k}$, at worst of scale
$M^{-i}$. We do the same in any non-planar ``primitive'' connected components (i.e. not containing sub non-planar components). The corresponding improvements are independant. \\

If a node of the Gallavotti-Nicol\`o tree is planar but has more than one
broken face ($B\ges 2$), we consider its number of external
legs\footnote{\label{fn:N2B2}It has been noticed in \cite{Chepelev2000hm} that
  orientable graphs can't have $N=2$ and $B=2$. A simple argument on the Filk
  rosette \cite{xphi4-05,Filk1996dm} proves it equally.}. If $N(G^{i}_{k})\ges
6$, we directly use lemma \ref{lem:masselottes}. When $N(G^{i}_{k})= 4$, the
number of broken faces is $1$ or $2$. Let us focus on the $B=2$ case. At scale
$i$, one or several lines contract above two external points $x'$ and $y'$.
In contrast with commutative field theory, the power counting of this
connected component depends on the scales down to $0$. Let $\scP$ the unique path in the Gallavotti-Nicol\`o tree linking $G^{i}_{k}$ to $G$. If there exists a scale $i_{0}<i$ and a connected component $G^{i_{0}}_{k'}$ on $\scP$ such that $N(G^{i_{0}}_{k'})=2$ then there exists lines of scales between $i$ and $i_{0}$ joining $x'$ to $y'$. Let us call $I$ the set of such lines and $i_{m-1}$ the scale of the first node after $G^{i}_{k}$ on $\scP$. If $\card I=1$ then $G^{i}_{k}$ is logarithmically divergent. If $\card I\ges 2$ then $G^{i}_{k}$ will be convergent as $M^{-2(i-i_{m-1})}$. Finally if there does not exist such a $G^{i_{0}}_{k'}$ then $G^{i}_{k}$ will be convergent as $M^{-2(i-i_{m-1})}$.\\

\begin{figure}[htbp]
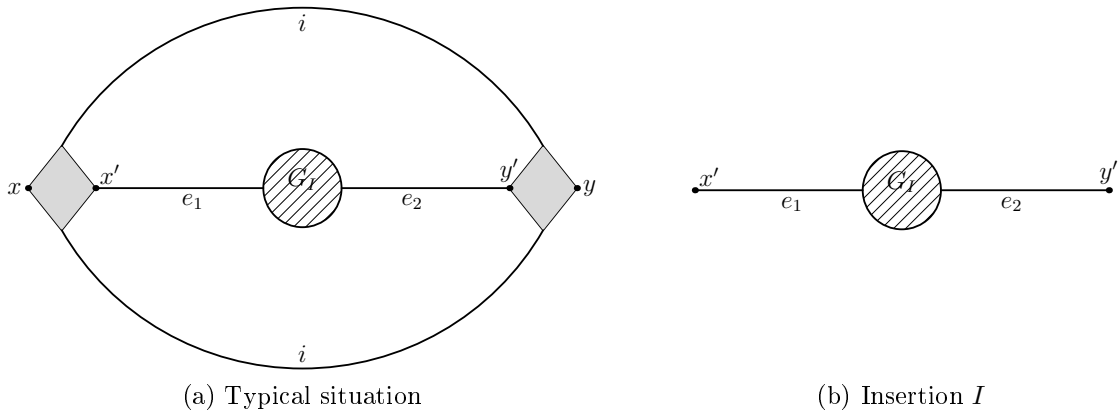

  \centering 
  \subfloat[Typical
  situation]{{\label{fig:newG-typ}}\includegraphics[scale=.8]{figures.3}}\qquad
  \subfloat[Insertion $I$]{\label{fig:insertion}\includegraphics[scale=.8]{figures.44}}
  \caption{Connected component (potentially) \emph{critical}}
  \label{fig:newG}
\end{figure}

Let us look at the figure \ref{fig:newG} which is simpler than the general situation but exhibits
all its important features. We define $I$ as the insertion made of the lines $e_{1}$,
$e_{2}$ and of the graph $G_{I}$. Note that $I$ may be empty and $G_{I}$ non-planar. The different
scales entering $I$ are $i_{0}<i_{1},\dotsc,i_{m-1}\,(<i_{m}=i)$. The corresponding connected component at scale $i_{0}$ is written $G^{i_{0}}_{k'}$. We also write $\cL_{I}$ for the set of loop lines in the insertion $I$.

\medskip
\noindent
We first get all the scaled decreasing functions for all the tree variables $v_{l}$ and $p_{l}$ except the lowest tree line $t$ in $I$. Then,
down to scale $i$, we proceed for the loop masslets as we have done for lemma \ref{lem:masselottes}. The total oscillation may be written
\begin{align}
  \varphi'_{\Omega}=&\varphi_{E}+\varphi_{X}+\sum_{\cT\setminus\{t\}}\veps(l)\cV_{l}\wed (u_{l}-\veps(l)a_{l})+\frac 12(1+\epsilon(t)\Omega)\veps(t)v_{t}\wed u_{t}\\
    &\hspace{-.5cm}+\sum_{\cT\setminus\{t\}}(1+\epsilon(l)\Omega)^{-1}\veps(l)\cV_{l}\!\cdot\!p_{l}+\sum_{\cL^i_{k}}\veps(\ell)\cW_{\ell}\wed
  (u_{\ell}-\veps(\ell)a_{\ell})+W'R_{1}P+PR_{2}P+PR_{3}U\notag\\
    &+\frac 12\sum_{\cL_{I}}(1+\epsilon(\ell)\Omega)\veps(\ell)w_{\ell}\wed u_{\ell}+\frac 12\sum_{\cL_{I}\ltimes\cL_{I}}\veps(\ell)w_{\ell}\wed u_{\ell'}+\veps(\ell')w_{\ell'}\wed u_{\ell}\nonumber\\
    &+\sum_{(\cL_{I}\cup \{t\})\subset\cL^i_{k}}\cW_{k}^{i}\wed u_{l}+\sum_{(\cL_{I}\cup \{t\})\subset\cL_{I}}\veps(\ell')w_{\ell'}\wed u_{\ell}
    +\sum_{k\subset\cL^i_{k}}\cW^{i}_{k}\wed\eta(k)x_{k}+\cW_{I}Q_{W\!X}X\notag\\
    &+\cW_{I}Q_{W}\cW_{I}+\sum_{\substack{(\cT\cup\cL)\prec(\cT\cup\cL)}}u_{l'}\wed u_{l}+\frac
    12\sum_{\substack{\cL_{I}\ltimes\cL_{I}}}u_{\ell'}\wed u_{\ell}+AR_{4}A+AR_{5}U+AR_{6}X\notag
\end{align}
where we wrote $\cW_{I}$ ($\cW^{i}_{k}$) for a linear combination of $\cW_{\ell},\,\ell\in\cL_{I}\,(\cL^{i}_{k})$. Let us pick one $\cW_{\ell},\,\ell\in\cL^{i}_{k}$. We use the oscillation $\cW_{\ell}\wed\big(\sum_{(\cL_{I}\cup \{t\})\subset\ell}u_{l}+\sum_{k\subset\ell}\eta(k)x_{k}\big)$ to get a decreasing function $\ks$ implementing $\big|\sum_{\cL_{I}\cup \{t\}}u_{l}+\sum_{k\subset\ell}\eta(k)x_{k}\big|\les M^{-i}$.

If there are external points overflown by the line $\ell$, there exists $k$ such that $x_{k}\defi z\subset\ell$. Then for all line in $\cL_{I}\cup\{t\}$, we perform the change of variables \eqref{eq:chgtvarVl} and \eqref{eq:chgtvarWl} but with $z$ in place of $x_{1}$. These modifications let  the function $\ks$ independant the $a_{l},\, l\in\cL_{I}\cup\{t\}$. This allows to get  for all line $l\in\cL_{I}\cup\{t\}$ a masslet of index $i_{l}$.

If there are no external point apart from $x$ and $y$ in $G^{i_{0}}_{k'}$ (see figure \ref{fig:newG}), the function $\ks$ only depends on $\sum_{(\cL_{I}\cup \{t\})\subset\ell}u_{l}$ and  $G^{i_{0}}_{k'}$ is a two-point graph. Let us write $\ell_{0}$ for the lowest line in $I$, $i_{\ell_{0}}=i_{0}$. Note that it is necessarily a loop line. For all line $\ell\in\cL_{I}\setminus\{\ell_{0}\}$, we perform
\begin{align}
  \lb
  \begin{aligned}
    u_{\ell}\to& u_{\ell}-\veps(\ell)a_{\ell},\\
    w_{\ell}\to& w_{\ell}+a_{\ell},\\
    u_{\ell_{0}}\to& u_{\ell_{0}}+\veps(\ell)a_{\ell},\\
    w_{\ell_{0}}\to& w_{\ell}-\veps(\ell_{0})\veps(\ell)a_{\ell}.
  \end{aligned}\right.\label{eq:chgtvarWl-I}
\end{align}
This let  $u_{\ell}+u_{\ell_{0}}$ (and $\ks$) fixed. Then for all line 
$\ell\in\cL_{I}\setminus\{\ell_{0}\}$, we get a decreasing function in
$\cW_{\ell}-\veps(\ell)\veps(\ell_{0})\cW_{\ell_{0}}$ of index $i_{\ell}$. All these functions are independant. For $\ell_{0}$, we perform
\begin{align}
    \lb
  \begin{aligned}
    u_{\ell_{0}}\to& u_{\ell_{0}}-\veps(\ell_{0})a_{\ell_{0}},\\
    w_{\ell_{0}}\to& w_{\ell_{0}}+a_{\ell_{0}},\\
    u_{t}\to&u_{t}+\veps(\ell_{0})a_{\ell_{0}}.
  \end{aligned}\right.\label{eq:chgtvarVl-I}
\end{align}
We get a decreasing function allowing to integrate over $\cW_{\ell_{0}}$ at the cost of $M^{i_{t}}\ges M^{i_{0}}$. Finally for the tree line $t$, we use the usual change of variables \eqref{eq:chgtvarVl}. This introduces $a_{t}$ in $\ks$. The masslet we get for $\cV_{t}$ is then of order $M^{i}$. Fortunately the corresponding strong decrease for $p_{t}$ is of order $M^{-i}$. We recover the fact that the long tree line variables do not cost anything.\\

\noindent
Let us call \textbf{critical} a four-point connected component with $N=4, g=0, B=2$ and the insertion $I$ reduced to a single line.
We are now ready to prove the following lemma
\begin{lemma}[Power counting]\label{lem:compt-puiss}
  Let $G$ an orientable connected graph. For all $\Omega\in\lsb 0,1\rbt$, there exists $K\in\R$ such that its
  amputed amplitude $A_{G}^{\mu}$ integrated over test functions (see
  (\ref{eq:amplitude-avt-massel})) is bounded by
  \begin{align}
    \labs A_{G}^{\mu}\rabs\les&K^{n}\prod_{i,k}M^{-\frac 12\omega(G^{i}_{k})}\label{eq:compt-bound}\\
    \text{with } \omega(G^{i}_{k})=&
    \begin{cases}
      N-4&\text{if ($N=2$ or $N\ges 6$) and $g=0$,}\\
      &\text{if $N=4$, $g=0$ and $B=1$,}\\
      &\text{if $G^{i}_{k}$ is critical,}\\
      N&\text{if $N=4$, $g=0$, $B=2$ and $G^{i}_{k}$ non-critical,}\\
      N+4&\text{if $g\ges 1$.}
    \end{cases}
  \end{align}
\end{lemma}
\begin{rem}
  This bound is not optimal but sufficient to prove the perturbative renormalizability of the theory. After the study of the propagator in the matrix basis \cite{toolbox05}, we could get the true power counting in particular the genus dependance. Concerning the broken faces, the bound \eqref{eq:compt-bound} is almost optimal. For the four-point function, it is. But for six (or more)-point functions, we did not try to improve our bound. Nevertheless remark that for such functions, similar situations to the four-point one may happen. The ``external'' points in additionnal broken faces may be linked by only one lower line. In this situation, the broken faces do not improve the power counting even for six (or more)-point functions. This is one of the differences between the Gross-Neveu model and the $\Phi^4$'s one.
\end{rem}
\begin{proof}
  Lemma \ref{lem:masselottes} allows to bound the amplitude of a connected orientable graph $G$ by
  \begin{align}
    \labs A^{\mu}_{G}\rabs\les&K^{n}\int
    dx_{1}\,g_{1}(x_{1}+\{a\})\delta_{G}\prod_{i=2}^{N}dx_{i}\,g_{i}(x_{i})\prod_{l\in G}da_{l}\,M^{2i_{l}}\Xi(a_{l})\label{eq:absbound-start}\\
    &\qquad\prod_{l\in\cT}du_{l}d\cV_{l}dp_{l}\,M^{i_{l}}e^{-M^{2i_{l}}(u_{l}-\veps(l)a_{l})^{2}}\prod_{\mu=0}^{1}\frac{1}{1+M^{-2i_{l}}\cV^{2}_{l,\mu}}
    \frac{1}{1+M^{2i_{l}}p^{2}_{l,\mu}}\notag\\
    &\qquad\prod_{\ell\in\cL}
    du_{\ell}d\cW_{\ell}M^{i_{\ell}}M^{i_{\ell}}
    e^{-M^{2i_{\ell}}(u_{\ell}+\{a\})^{2}}
    \prod_{\mu=0}^{1}\frac{1}{1+M^{-2i_{\ell}}\cW^{2}_{\ell,\mu}}\notag
  \end{align}
  where $K\in\R$ and $g_{i},\,i\in\lnat 1,N\rnat$ and $\Xi$ are Schwartz-class functions. The $\delta_{G}$ function corresponding to the root delta function is given by (see section \ref{sec:resol-delta})
  \begin{equation}
    \delta_{G}\Big(\sum_{i\in\cE(G)}\eta(i)x_{i}+\sum_{l\in\cT\cup\cL}u_{l}\Big).
  \end{equation}
  We use it to integrate over one of the external positions. The other ones are integrated with the $g_{i}$'s functions. The bound (\ref{eq:absbound-start}) on the absolute value of the amplitude becomes
    \begin{align}
    \labs A^{\mu}_{G}\rabs\les&K^{n}\int
    \prod_{l\in G}da_{l}\,M^{2i_{l}}\Xi(a_{l})\prod_{\ell\in\cL}
    du_{\ell}d\cW_{\ell}M^{i_{\ell}}M^{i_{\ell}}
    e^{-M^{2i_{\ell}}(u_{\ell}+\{a\})^{2}}
    \prod_{\mu=0}^{1}\frac{1}{1+M^{-2i_{\ell}}\cW^{2}_{\ell,\mu}}\label{eq:absbound-2}\\
    &\qquad\prod_{l\in\cT}du_{l}d\cV_{l}dp_{l}\,M^{i_{l}}e^{-M^{2i_{l}}(u_{l}-\veps(l)a_{l})^{2}}\prod_{\mu=0}^{1}\frac{1}{1+M^{-2i_{l}}\cV^{2}_{l,\mu}}
    \frac{1}{1+M^{2i_{l}}p^{2}_{l,\mu}}.\notag
  \end{align}
  The integrations over the $a_{\ell}$ variables cost $\cO(1)$. For all line $l$ in the graph, integration over $u_{l}$ is of order $\cO(M^{-2i_{l}})$. The integration over $v_{l}$ (resp. $w_{l}$) is of order $\cO(M^{2i_{l}})$. But for tree lines, this is compensated by the integration over $p_{l}$ which gives $\cO(M^{-2i_{l}})$. Then the loops only cost $\cO(1)$ whereas the tree lines earn $\cO(M^{-2i_{l}})$. We have the following bound
  \begin{align}
    \labs A_{G}^{\mu}\rabs\les& K^{n}\prod_{l\in
      G}M^{i_{l}}\prod_{l\in\cT}M^{-2i_{l}}\notag\\
    \les&K'^{n}\prod_{l\in
      G}M^{i_{l}+1}\prod_{l\in\cT}M^{-2(i_{l}+1)}.
  \end{align}
  We may now distribute the power counting among the connected components \cite{Riv1} :
  \begin{align}
    \prod_{l\in G}M^{i_{l}+1}=&\prod_{l\in G}\prod_{i=0}^{i_{l}}M=\prod_{l\in
      G}\prod_{\substack{(i,k)\in\N^{2}/\\l\in
        G^{i}_{k}}}M=\prod_{(i,k)\in\N^{2}}\,\prod_{l\in G^{i}_{k}}M,\\
    \prod_{l\in\cT}M^{-2(i_{l}+1)}=&\prod_{l\in\cT}\prod_{\substack{(i,k)\in\N^{2}/\\l\in
        G^{i}_{k}}}M^{-2}=\prod_{(i,k)\in\N^{2}}\,\prod_{l\in\cT^{i}_{k}}M^{-2}.
  \end{align}
Then, changing $K'$ into $K$, the amplitude of a connected orientable graph is bounded by
\begin{align}
  \labs A_{G}^{\mu}\rabs\les&K^{n(G)}\prod_{(i,k)\in\N^{2}}M^{-\frac
    12\omega(G^{i}_{k})},\\
  \text{where }\omega(G^{i}_{k})=&N(G^{i}_{k})-4\label{eq:degre-conv-compconn}
\end{align}
which proves the first part of lemma \ref{lem:compt-puiss}.\\

If a connected component $G^{i}_{k}$ is non-planar, there exists $\ell,\ell'\in
G^{i}_{k}$ such that the integration over $\cW_{\ell}$ gives
$M^{-2i_{\ell'}}\les M^{-2i}$ instead of $M^{2i_{\ell}}$ (see section \ref{subsec:nonplanar}). The gain with respect to \eqref{eq:degre-conv-compconn} is at least $M^{-4i}$. The superficial degree of convergence becomes $\omega(G^{i}_{k})=N(G^{i}_{k})+4$.\\

Finally let a connected component $G^{i}_{k}$ with four external legs and two broken faces. With the notations previously defined, if $G^{i_{0}}_{k'}$ has more than two external points, we use the function $\ks$ to integrate over one of these external positions. This brings $M^{-2i}$ instead of $\cO(1)$. Let us write $\mathbf{\scP}$ for the path in the Gallavotti-Nicolò tree between $G^i_{k}$ and $G$. The factor $M^{-2i}$ improves the superficial degree of convergence of all the nodes in $\scP$ with $N=4, B=2$. It becomes  $\omega(G^{i}_{k})=N(G^{i}_{k})$. If $G^{i_{0}}_{k'}$ is a two-point graph, we use $\ks$ to integrate over the $u$ variable of the lowest line in $I$. This brings $M^{-2i}$ instead of $M^{-2i_{0}}$. The gain with respect to \eqref{eq:degre-conv-compconn} is then $M^{-2(i-i_{0})}$. But the integration over $\cW_{\ell_{0}}$ costs $M^{2i_{t}}$ instead of $M^{2i_{\ell_{0}}}$. The total gain is then only $M^{-2(i-i_{t})}$. This additionnal factor allows to improve the power counting of all the four-point components with $B=2$ in $\scP$ between $G^{i}_{k}$ and the scale $i_{t}$. Their power counting increase from $N-4$ to $N$. But note that between $i_{t}$ (the scale of the lowest tree line in $I$) and $i_{0}$, only loop lines may appear in the subgraphs. Then the number of external points may only \emph{strictly} decrease in $\scP$ from scale $i_{t}$ to scale $i_{0}$. $G^{i_{0}}_{k'}$ being a two-point graph, there may be only one divergent connected component in $\scP$ between $i_{t}$ and $i_{0}$. It is a four-point graph with $B=2$ at scale $i_{1}$ (the lowest scale in $I$ above $i_{0}$). Moreover this happens only if there is only one loop line of scale $i_{0}$. This component is \emph{critical} (by definition) and we can't improve its power counting which remains $N-4$. This proves lemma \ref{lem:compt-puiss}.
\end{proof}

\section{Renormalization}
\label{sec:renorm-GN}

Thanks to the power counting proved in lemma \ref{lem:compt-puiss}, we
know that the only divergent subgraphs are the planar two- and
four-point ones. More precisely the only divergent two-point graphs
have one broken face. The divergent four-point ones have either
one broken face or are \emph{critical} which means they have
$N=4,g=0,B=2$ and the two ``external'' points belonging to the second
broken face are linked by one (and only one) line of lower scale.
We are going to prove that the divergent parts of those graphs are of the form of the initial Lagrangian.

\subsection{The four-point function}
\label{subsec:4pt-fct}

\subsubsection{$B=1$}

Let a planar four-point subgraph with one broken face needing renormalization. It is then a node of the Gallavotti-Nicol\`o tree. There exists $(i,k)\in\N^{2}$ such that $N(G^{i}_{k})=4,g(G^{i}_{k})=0,B(G^{i}_{k})=1$. The four external points of this amputed graph are written $x_{j},\,j\in\lnat 1,4\rnat$. The amplitude associated to the connected component $G^{i}_{k}$ is 
\begin{align}
  A^{\mu}_{G^{i}_{k}}(\{x_{j}\})=\int&\prod_{i=1}^{4}dx_{i}\,\psib_{e}(x_{1})\psi_{e}(x_{2})\psib_{e}(x_{3})
  \psi_{e}(x_{4})\delta_{G^{i}_{k}} e^{\imath\varphi'_{\Omega}}\\
  &\prod_{l\in\cT^{i}_{k}}du_{l}dv_{l}dp_{l}\,\bar{C}^{i_{l}}_{l}(u_{l})\prod_{\ell\in\cL^{i}_{k}}du_{\ell}dw_{\ell}\,
  \bar{C}^{i_{\ell}}_{\ell}(u_{\ell})\notag
\end{align}
where $e$ is the biggest external index of the subgraph $G^{i}_{k}$ and
$\psi_{e},\psib_{e}$ are fields of indices lower or equal to $e<i$. We will perform a first order Taylor expansion which will allow to decouple the external variables $x_{j}$ from the internal ones $u$ and $p$ and identify the divergent part of the amplitude. We introduce a parameter $s$ in three different places. First of all, we expand the delta function $\delta_{G^{i}_{k}}$ as
\begin{align}
  \delta_{G^{i}_{k}}\big(\Delta+s\kU\big)\Big|_{s=1}=&\delta(\Delta)+\int_{0}^{1}
  ds\,\kU\cdot\nabla\delta(\Delta+s\kU)\\
                                %\notag\\
  \text{where }\Delta=&x_{1}-x_{2}+x_{3}-x_{4}\text{ and }\kU=\sum_{l\in G^{i}_{k}}u_{l}.\notag
\end{align}
For orientable graphs, the fields $\psib$ are associated to odd positions and the $\psi$'s to even ones. Moreover if the graph is planar regular, corollary \ref{sec:oscillRG} gives the exact value of the root delta function, in particular the alternating signs. This corollary also gives the external oscillation $\varphi_{E}$. The remaining oscillation $\varphi'_{\Omega}$ is now expanded. It is given by corollary \ref{sec:oscillRG} and by the branch delta functions oscillations. With (hopefully) self-explaining notations, it may be written
\begin{align}
  \varphi'_{\Omega}(s=1)=\varphi_{E}+XQ_{X\!U}U+XQ_{X\!P}P+UQ_{U}U+PQ_{P}P+UQ_{U\!W}W+PQ_{PW}W.
\end{align}
Remark that $Q_{X\!W}=Q_{W}=0$ for planar regular graphs. We write
\begin{align}
  &\exp\imath(XQ_{X\!U}U+XQ_{X\!P}P+UQ_{U}U+PQ_{P}P)\\
  =&1+\imath\int_{0}^{1}ds\,(XQ_{X\!U}U+XQ_{X\!P}P+UQ_{U}U+PQ_{P}P)e^{\imath s(XQ_{X\!U}U+XQ_{X\!P}P)+\imath UQ_{U}U+\imath PQ_{P}P}.\notag
\end{align}
Finally we also expand the internal propagators. For all line $l\in G^{i}_{k}$,
\begin{align}
  \bar{C}_{l}(u_{l},s=1)=&\frac{\Omega}{\theta\pi}\int_{0}^{\infty}\frac{dt_{l}\,e^{-t_{l}m^{2}}}{\sinh(2\Ot t_{l})}\,  
  e^{-\frac{\Ot}{2}\coth(2\Ot t_{l})u_{l}^{2}}\big(\imath\Ot\coth(2\Ot
  t_{l})\epsilon(l)\veps(l)\us_{l}\label{eq:taylor-propa4}\\
&\qquad+s\Omega\epsilon(l)\veps(l)\uts_{l}+sm\big)\big(\cosh(2\Ot t_{l})\mathds{1}_{2}-s\imath{\textstyle\frac{\theta}{2}}\sinh(2\Ot
  t_{l})\gamma\Theta^{-1}\gamma\big)\Big|_{s=1}\notag\\
  =&\frac{2\imath \Omega^{2}}{\theta^{2}\pi}\int_{0}^{\infty}\frac{dt_{l}\,e^{-t_{l}m^{2}}}{\tanh(2\Ot t_{l})}\,  
  e^{-\frac{\Ot}{2}\coth(2\Ot t_{l})u_{l}^{2}}\coth(2\Ot
  t_{l})\epsilon(l)\veps(l)\us_{l}\notag\\
  &+\frac{\Omega}{\theta\pi}\int_{0}^{1}ds\,\int_{0}^{\infty}\frac{dt_{l}\,
    e^{-t_{l}m^{2}}}{\sinh(2\Ot t_{l})}\,e^{-\frac{\Ot}{2}\coth(2\Ot
    t_{l})u_{l}^{2}}\notag\\
  &\qquad\times\Big\{(\Omega\epsilon(l)\veps(l)\uts_{l}+m)\big(\cosh(2\Ot
  t_{l})\mathds{1}_{2}-s\imath{\textstyle\frac{\theta}{2}}\sinh(2\Ot
  t_{l})\gamma\Theta^{-1}\gamma\big)\notag\\ 
  &\qquad-\imath{\textstyle\frac{\theta}{2}}\sinh(2\Ot t_{l})\big(\imath\Ot\coth(2\Ot
  t_{l})\epsilon(l)\veps(l)\us_{l}+s\Omega\epsilon(l)\veps(l)\uts_{l}+sm\big)\gamma\Theta^{-1}\gamma\Big\}.\notag
\end{align}
Let $\tau A$ be the counterterm associated to the connected component $G^{i}_{k}$. It corresponds
to the zeroth order terms of the three preceding expansions :
\begin{align}
  \tau A^{\mu}_{G^{i}_{k}}=&\int\prod_{i=1}^{4}dx_{i}\,\psib_{e}(x_{1})\psi_{e}(x_{2})\psib_{e}(x_{3})
  \psi_{e}(x_{4})\delta(\Delta) e^{\imath\varphi_{E}}\\
  &\times\int\prod_{l\in\cT^{i}_{k}}du_{l}dv_{l}dp_{l}\,\bar{C}^{i_{l}}_{l}(u_{l},s=0)
  \prod_{\ell\in\cL^{i}_{k}}du_{\ell}dw_{\ell}\,
  \bar{C}^{i_{\ell}}_{\ell}(u_{\ell},s=0)e^{\imath\varphi'_{\Omega}(s=0)}\notag
  \intertext{where $\varphi_{E}=\sum_{i<j=1}^{4}(-1)^{i+j+1}x_{i}\wed
    x_{j}$. Then the counterterm is of the form}
  \tau
  A^{\mu}_{G^{i}_{k}}=&\int dx\,\lbt\psib_{e}\star\psi_{e}\star\psib_{e}\star\psi_{e}\rbt(x)\\
  &\times\int\prod_{l\in\cT^{i}_{k}}du_{l}dv_{l}dp_{l}\,\bar{C}^{i_{l}}_{l}(u_{l},s=0)
  \prod_{\ell\in\cL^{i}_{k}}du_{\ell}dw_{\ell}\,
  \bar{C}^{i_{\ell}}_{\ell}(u_{\ell},s=0)e^{\imath\varphi'_{\Omega}(s=0)}.\notag
\end{align}
To prove that $\tau A$ looks like the initial vertex, it remains to show that its spinorial structure is one of those of equation \eqref{eq:int-orient}. Apart from the oscillations and the exponential decreases of the propagators, the counterterm $\tau A$ involves
\begin{align}\label{eq:polyn}
  P=&\prod_{l\in G}\us_{l}=\prod_{l\in G}\lbt\gamma^{0}u_{l}^{0}+\gamma^{1}u_{l}^{1}\rbt=\prod_{i=1}^{2n-N/2}P_{i}.
\end{align}
Each of the $2^{2n-N/2}$ terms $P_{i}$ in $P$ consist in choosing for each line $l\in G$ either $\gamma^{0}u^{0}_{l}$ or $\gamma^{1}u_{l}^{1}$.  Each $P_{i}$ has $n_{i}^{0}$ $u^{0}$ and $n_{i}^{1}$ $u^{1}$. Note that, apart from $P$, the counterterm $\tau A$ is invariant under: 
$\forall l\in G,\, u_{l}^{0}\to -u_{l}^{0}$ and $w_{l}^{1}\to -w_{l}^{1}$. Then the only non vanishing $P_{i}$ have even $n_{i}^{0}$. With a similar argument we prove that $n_{i}^{1}$ is also even. Each term in $\tau A$ then consists in even numbers of $\gamma^{0}$ and $\gamma^{1}$. For the four-point function, the Taylor expansion \eqref{eq:taylor-propa4} is possible because the number of internal lines is even (it is $2(n-1)$). We now define the notions of \textbf{chain} and \textbf{cycle}. 
\begin{defn}[Chain and cycle]\label{def:cycle-chaine}
  We say that two fields are in the same chain
  \begin{itemize}
  \item if they both belong to a same scalar product at a vertex\footnote{For example, the first two fields in the interaction \eqref{eq:int-o-1}
belong to a same scalar product.},
  \item if they are linked by a propagator.
  \end{itemize}
A cycle is a closed chain.
\end{defn}
The external fields are linked by chains. The other (internal) fields belong to cycles. The $\gamma^{0}$ and $\gamma^{1}$ matrices are distributed among chains and cycles. Each cycle corresponds, up to a sign, to a term $\Tr\lbt(\gamma^{0})^{p}(\gamma^{1})^{q}\rbt$. It does not vanish only if $p$ and $q$ are even. Knowing that the total number of $\gamma^{0}$ is 
even, that the total number of $\gamma^{1}$ is even and that each cycle contains even numbers of $\gamma^{0}$ and $\gamma^{1}$, the chains of the graph share an even number of $\gamma^{0}$ and an even number of $\gamma^{1}$. There are two chains in the four-point function graphs. There are then four possibilities to distribute the gamma matrices among these two chains. Each may contain an even or an odd number of $\gamma^{0}$ or $\gamma^{1}$.\\

Depending on the number and the type of the vertices in these two chains, they may link either a $\psi$ to a $\psib$ or two fields of the same kind. We are faced to twelve different spinorial structures:
\begin{subequations}
  \begin{align}
    \psi\star\lsb\lbt\gamma^{0}\rbt^{2p}\lbt\gamma^{1}\rbt^{2q}\rsb\psib\star
    \psi\star\lsb\lbt\gamma^{0}\rbt^{2p'}\lbt\gamma^{1}\rbt^{2q'}\rsb\psib
    =&\pm\psi\star\mathds{1}\psib\star\psi\star\mathds{1}\psib,\\
    &\notag\\
    \psi\star\lsb\lbt\gamma^{0}\rbt^{2p+1}\lbt\gamma^{1}\rbt^{2q}\rsb\psib\star
    \psi\star\lsb\lbt\gamma^{0}\rbt^{2p'+1}\lbt\gamma^{1}\rbt^{2q'}\rsb\psib
    =&\pm\psi\star\gamma^{0}\psib\star\psi\star\gamma^{0}\psib,\\
    &\notag\\
    \psi\star\lsb\lbt\gamma^{0}\rbt^{2p}\lbt\gamma^{1}\rbt^{2q+1}\rsb\psib\star
    \psi\star\lsb\lbt\gamma^{0}\rbt^{2p'}\lbt\gamma^{1}\rbt^{2q'+1}\rsb\psib
    =&\pm\psi\star\gamma^{1}\psib\star\psi\star\gamma^{1}\psib,\\
    &\notag\\
    \psi\star\lsb\lbt\gamma^{0}\rbt^{2p+1}\lbt\gamma^{1}\rbt^{2q+1}\rsb\psib\star
    \psi\star\lsb\lbt\gamma^{0}\rbt^{2p'+1}\lbt\gamma^{1}\rbt^{2q'+1}\rsb\psib
    =&\pm\psi\star\gamma^{0}\gamma^{1}\psib\star\psi\star\gamma^{0}\gamma^{1}\psib.
  \end{align}
\end{subequations}
In the same way, we can meet
\begin{subequations}
  \begin{align}
    \pm\psib\star\mathds{1}\psi\star\psib\star\mathds{1}\psi,\\
    &\notag\\
    \pm\psib\star\gamma^{0}\psi\star\psib\star\gamma^{0}\psi,\\
    &\notag\\
    \pm\psib\star\gamma^{1}\psi\star\psib\star\gamma^{1}\psi,\\
    &\notag\\
    \pm\psib\star\gamma^{0}\gamma^{1}\psi\star\psib\star\gamma^{0}\gamma^{1}\psi,
  \end{align}
\end{subequations}
\begin{subequations}
  \begin{align}
    \pm\psib_{a}\star\psi_{c}\star\psib_{b}\star\psi_{d}\mathds{1}_{ab}\mathds{1}_{cd},\\
    &\notag\\
    \pm\psib_{a}\star\psi_{c}\star\psib_{b}\star\psi_{d}\gamma^{0}_{ab}\gamma^{0}_{cd},\\
    &\notag\\
    \pm\psib_{a}\star\psi_{c}\star\psib_{b}\star\psi_{d}\gamma^{1}_{ab}\gamma^{1}_{cd},\\
    &\notag\\
    \pm\psib_{a}\star\psi_{c}\star\psib_{b}\star\psi_{d}\lbt\gamma^{0}\gamma^{1}
    \rbt_{ab}\lbt\gamma^{0}\gamma^{1}\rbt_{cd}.
  \end{align}
\end{subequations}
To prove that the divergence of the four-point function is of the form of the original vertices \eqref{eq:int-orient}, it is convenient to rewrite them in a different way.

\paragraph{\Nc{} Fierz identities}

A basis for $M_{D}(\C)$ is given by a representation of the Clifford algebra $\lb\gamma^{\mu},\gamma^{\nu}\rb=-D\delta^{\mu\nu}$ of dimension $D$. In dimension $2$, $\cB=\lb\Gamma^{0}=\mathds{1},\Gamma^{1}=\gamma^{0},\Gamma^{2}=\gamma^{1},\Gamma^{3}=\gamma^{0}\gamma^{1}\rb$ is a basis for $M_{2}(\C)$. Then let $M\in M_{2}(\C)$,
\begin{align}\label{eq:cliff-decomp}
  M=&-\frac 12\sum_{A,B=0}^{3}\eta_{AB}\Tr(M\Gamma^{A})\Gamma^{B},\\
  \text{with }\eta=&\diag(-1,1,1,1).\notag
\end{align}
We now use such a decomposition to rewrite the interactions of the model under a different form. For example, let us consider interaction \eqref{eq:int-o-2}. If we define $M_{ab}=\psib_{b}\star\psi_{a}$ and use \eqref{eq:cliff-decomp}, we have
\begin{align}
  \psib_{b}\star\psi_{a}=&-\frac 12\sum_{A,B}\eta_{AB}\psib_{b'}\star\psi_{a'}\Gamma^{A}_{b'a'}\Gamma^{B}_{ab}.
\end{align}
This allows to write
\begin{align}
  \int\psi_{a}\star\psib_{a}\star\psi_{b}\star\psib_{b}=\int\psib_{b}\star\psi_{a}\star\psib_{a}\star\psi_{b}
  =-\frac
  12\sum_{A,B}\eta_{AB}\int\psib\star\Gamma^{A}\psi\star\psib\star\Gamma^{B}\psi.
  %% =&-\frac
%%   12\sum_{A,B}\delta_{AB}\int\psib\star\Gamma^{A}\psi\star\psib\star\Gamma^{B}\psi.\notag
\end{align}
In the same way, for interaction \eqref{eq:int-o-3}, we use the decomposition 
\begin{align}
 M_{ba}=&\psib_{a}\star\psi_{b}=-\frac 12\sum_{A,B}\eta_{AB}\psib_{a'}\star\psi_{b'}\Gamma^{A}_{a'b'}\Gamma^{B}_{ba}
\end{align}
and write
\begin{align}
  \sum_{a,b}\int
  \psib_{a}\star\psi_{b}\star\psib_{a}\star\psi_{b}
  =&-\frac
  12\sum_{A,B}\eta_{AB}\int\psib\star\Gamma^{A}\psi\star\psib\star\!\!\!\phantom{\Gamma}^{t\!}\Gamma^{B}\psi\notag\\
  =&-\frac
  12\sum_{A,B}\int g^{3}_{AB}\psib\star\Gamma^{A}\psi\star\psib\star\Gamma^{B}\psi
\end{align}
with $g^{3}_{AB}=\diag(-1,1,1,-1)$. We do the same for the three other interactions. The six possible interactions are given in table\footnote{Remind that we restrict our proof to the orientable case.} \ref{tab:interactions}. As a conclusion, the three orientable interactions \eqref{eq:int-orient} may be written as linear combinations of
\begin{subequations}\label{eq:courants-or}
  \begin{align}
    &\int\psib\star\bbbone\psi\star\psib\star\bbbone\psi,\label{eq:courants-or-1}\\
    &\int\psib\star\gamma^{\mu}\psi\star\psib\star\gamma_{\mu}\psi\text{ and}\label{eq:courants-or-2}\\
    &\int\psib\star\gamma^{0}\gamma^{1}\psi\star\psib\star\gamma^{0}\gamma^{1}\psi\label{eq:courants-or-3}
  \end{align}
\end{subequations}
whereas the non-orientable ones \eqref{eq:int-nonorient} may be written in function of 
\begin{subequations}\label{eq:courants-no}
  \begin{align}
    &\int\psi\star\bbbone\psib\star\psib\star\bbbone\psi,\label{eq:courants-no-1}\\
    &\int\psi\star\gamma^{\mu}\psib\star\psib\star\gamma_{\mu}\psi\text{ and}\label{eq:courants-no-2}\\
    &\int\psi\star\gamma^{0}\gamma^{1}\psib\star\psib\star\gamma^{0}\gamma^{1}\psi.\label{eq:courants-no-3}
  \end{align}
\end{subequations}
In equations \eqref{eq:courants-or-2} and \eqref{eq:courants-no-2}, the sum over $\mu$ is implicit.

\begin{landscape}
  \begin{table}\label{tab:interactions}
    \centering
    \begin{equation*}
      \begin{array}{|l|l|}
        \hline
        \multicolumn{2}{|c|}{\rule[-3pt]{0pt}{20pt}\text{\Large \bfseries Interactions of the \nc{} Gross-Neveu model}}\\
        \hline
        \hline
        \multicolumn{1}{|c|}{\rule[-3pt]{0pt}{15pt}\text{\large
            Orientable}}&\multicolumn{1}{|c|}{\text{\large Non-orientable}}\\
        \hline
        \rule[5pt]{0pt}{15pt}{\displaystyle\bullet\phantom{=}\sum_{a,b}\int
        dx\,\lbt\psib_{a}\star\psi_{a}\star\psib_{b}\star\psi_{b}\rbt(x)\phantom{=\Tr(\psib\cdot\psi)^2}}&
        {\displaystyle\bullet\phantom{=}\sum_{a,b}\int
        dx\,\lbt\psib_{a}\star\psib_{b}\star\psi_{b}\star\psi_{a}\rbt(x)\phantom{=\Tr(\psib\cdot\psi)^2}}\\
        \rule[0pt]{0pt}{15pt}{\displaystyle\phantom{\bullet}=-\frac{1}{2}
        \sum_{A,B}\int g^{1}_{AB}\psib\star\Gamma^{A}\psi\star\psib\star\Gamma^{B}\psi}&
        {\displaystyle\phantom{\bullet}=-\frac{1}{2}
        \sum_{A,B}\int g^{1}_{AB}\psi\star\Gamma^{A}\psib\star\psib\star\Gamma^{B}\psi}\\
        &\\
        &\\
        {\displaystyle\rule[0pt]{0pt}{15pt}\bullet\phantom{=}\sum_{a,b}\int
        dx\,\lbt\psi_{a}\star\psib_{a}\star\psi_{b}\star\psib_{b}\rbt(x)}&
        {\displaystyle\bullet\phantom{=}\sum_{a,b}\int
        dx\,\lbt\psib_{a}\star\psib_{a}\star\psi_{b}\star\psi_{b}\rbt(x)}\\
        {\displaystyle\rule[0pt]{0pt}{15pt}\phantom{\bullet}=-\frac{1}{2}
        \sum_{A,B}\int g^{2}_{AB}\psib\star\Gamma^{A}\psi\star\psib\star\Gamma^{B}\psi}&
        {\displaystyle\phantom{\bullet}=-\frac{1}{2}
        \sum_{A,B}\int g^{2}_{AB}\psi\star\Gamma^{A}\psib\star\psib\star\Gamma^{B}\psi}\\
        &\\
        &\\
        {\displaystyle\rule[0pt]{0pt}{15pt}\bullet\phantom{=}\sum_{a,b}\int
        dx\,\lbt\psib_{a}\star\psi_{b}\star\psib_{a}\star\psi_{b}\rbt(x)}&
        {\displaystyle\bullet\phantom{=}\sum_{a,b}\int
        dx\,\lbt\psib_{a}\star\psib_{b}\star\psi_{a}\star\psi_{b}\rbt(x)}\\
        {\displaystyle\rule[-10pt]{0pt}{25pt}\phantom{\bullet}=-\frac{1}{2}
        \sum_{A,B}\int g^{3}_{AB}\psib\star\Gamma^{A}\psi\star\psib\star\Gamma^{B}\psi}&
        {\displaystyle\phantom{\bullet}=-\frac{1}{2}
        \sum_{A,B}\int
        g^{3}_{AB}\psi\star\Gamma^{A}\psib\star\psib\star\Gamma^{B}\psi}\\
        \hline
        \multicolumn{2}{|c|}{{\displaystyle\rule[2pt]{0pt}{15pt}g^{1}=\diag(-2,0,0,0),\ g^{2}=\eta=\diag(-1,1,1,1),
          \ g^{3}=\diag(-1,1,1,-1)}}\\
        \multicolumn{2}{|c|}{{\displaystyle\rule[-8pt]{0pt}{23pt}\forall A\in\lnat
            1,4\rnat,\,\Gamma^{A}\in\cB=\lb\Gamma^{0}=\mathds{1},\Gamma^{1}=\gamma^{0},\Gamma^{2}=\gamma^{1},
            \Gamma^{3}=\gamma^{0}\gamma^{1}\rb}}\\
          \hline
        \end{array}
      \end{equation*}
      \caption{The interactions and their different formulations}
    \end{table}
  \end{landscape}
We now show that for all $\Gamma^{C}\in\cB,\,\psi\star\Gamma^{C}\psib\star\psi\star\Gamma^{C}\psib$,
$\psib\star\Gamma^{C}\psi\star\psib\star\Gamma^{C}\psi$ and
$\psib_{a}\star\psi_{c}\star\psib_{b}\star\psi_{d}\Gamma^{C}_{ab}\Gamma^{C}_{cd}$
may be expressed in function of the orientable interactions of table \ref{tab:interactions} with the help of a symmetry of the model. 
\begin{align}
  \int\psi\star\Gamma^{C}\psib\star\psi\star\Gamma^{C}\psib=&\int\psib_{d}\star\psi_{a}\star\psib_{b}\star\psi_{c}
  \Gamma^{C}_{ab}\Gamma^{C}_{cd}\\
  =&-\frac
  12\sum_{A,B}\eta_{AB}\psib\star\Gamma^{A}\psi\star\psib\star\!\!\!\!\!\phantom{O}^{t\!}\Gamma^{C}
  \Gamma^{B}\!\!\!\!\phantom{O}^{t\!}\Gamma^{C}\psi\notag\\
  \phantom{O}^{t\!}\Gamma^{C}\Gamma^{B}\!\!\!\!\phantom{O}^{t\!}\Gamma^{C}=&
  \begin{cases}
    \Gamma^{B}&\text{if $\Gamma^{C}=\mathds{1}$}\\
    g_{BB'}\Gamma^{B'}&\text{with $g=\diag(-1,1,1-1)$ if
      $\Gamma^{C}=\gamma^{0}\gamma^{1}$}\\
    g_{BB'}\Gamma^{B'}&\text{with $g=\diag(-1,-1,1,1)$ if
      $\Gamma^{C}=\gamma^{0}$}\\
    g_{BB'}\Gamma^{B'}&\text{with $g=\diag(-1,1,-1,1)$ if
      $\Gamma^{C}=\gamma^{1}$}
  \end{cases}
\end{align}
Then we have
\begin{align}
  \int\psi\star\Gamma^{C}\psib\star\psi\star\Gamma^{C}\psib=&-\frac
    12\sum_{A,B} g_{AB}\psib\star\Gamma^{A}\psi\star\psib\star\Gamma^{B}\psi\label{eq:inter1}\\
    \text{with }g=&
  \begin{cases}
    \diag(-1,1,1,1)&\text{if }\Gamma^{C}=\mathds{1}\\
    \diag(1,1,1-1)&\text{if }\Gamma^{C}=\gamma^{0}\gamma^{1}\\
    \diag(1,-1,1,1)&\text{if }\Gamma^{C}=\gamma^{0}\\
    \diag(1,1,-1,1)&\text{if }\Gamma^{C}=\gamma^{1}.
  \end{cases}\label{eq:or2-contreterme}
\end{align}
If $\Gamma^{C}=\mathds{1}$ or $\gamma^{0}\gamma^{1}$, the interaction
\eqref{eq:inter1} may be written in function of the interactions \eqref{eq:courants-or}.
On the contrary, if $\Gamma^{C}=\gamma^{0}$ or $\gamma^{1}$ independently, it is impossible.
Fortunately there exists a symmetry implying that the counterterms associated to interaction \eqref{eq:inter1} for $\Gamma^{C}=\gamma^{0}$ and $\gamma^{1}$ are equal. Each term $P_{i}$ in the polynomial $P$ \eqref{eq:polyn} consists indeed to choose, for each line in the graph, either $\gamma^{0}$ or $\gamma^{1}$. To each of these terms is canonically associated an other term $\bar{P}_{i}=P_{j},\,j\neq i$ for which we have done exactly the inverse choice. Then to get $\bar{P}_{i}$, we consider $P_{i}$ and change $\gamma^{0}$
into $\gamma^{1}$, $u_{l}^{0}$ into $u_{l}^{1}$ and vice-versa. Each counterterm, associated to a $P_{i}$, is made of a product of gamma matrices and of integrals over the variables $u_{l}$, $p_{l}$, $v_{l}$ and $w_{l}$. The rotation
\begin{align}
 \forall l\in G,\,u_{l}^{0}\to& u_{l}^{1}\\
 u_{l}^{1}\to& -u_{l}^{0}\notag\\
 w_{l}^{0}\,(v_{l}^0)\to& w_{l}^{1}\,(v_{l}^1)\notag\\
 w_{l}^{1}\,(v_{l}^1)\to& -w_{l}^{0}\,(-v_{l}^0)\notag
\end{align}
shows that the integrals in $\bar{P}_{i}$ equals the ones in $P_{i}$ (the total number of $u_{l}^{1}$ is even). Let us have a look at the products of gamma matrices. Let $N\in\N$ and $\forall j\in\lnat 0,2N+1\rnat,\,n_{j}\in\N$. 
\begin{align}
  P_{\gamma}=&\prod_{i=0}^{N}\lbt\gamma^{0}\rbt^{n_{2i}}\lbt\gamma^{1}\rbt^{n_{2i+1}}\notag\\
  =&\prod_{i=0}^{N}(-1)^{\lsb\frac{n_{2i}}{2}\rsb+\lsb\frac{n_{2i+1}}{2}\rsb}\lbt\gamma^{0}
  \rbt^{\frac{1-(-1)^{n_{2i}}}{2}}\lbt\gamma^{1}\rbt^{\frac{1-(-1)^{n_{2i+1}}}{2}}.\label{eq:pgamma}
\end{align}
Each product of $\gamma^{0}$ (resp. $\gamma^{1}$) has been reduced thanks to $\lbt\gamma^{0}\rbt^{2}=\lbt\gamma^{1}\rbt^{2}=-\mathds{1}$. The product $P_{\gamma}$ equals, up to a sign, an \emph{alternating} product $P_{\gamma}^{\text{a}}$ of $\gamma^{0}$ and $\gamma^{1}$. In the same way,
\begin{align}
  \bar{P}_{\gamma}=&\prod_{i=0}^{N}\lbt\gamma^{1}\rbt^{n_{2i}}\lbt\gamma^{0}\rbt^{n_{2i+1}}\notag\\
  =&\prod_{i=0}^{N}(-1)^{\lsb\frac{n_{2i}}{2}\rsb+\lsb\frac{n_{2i+1}}{2}\rsb}\lbt\gamma^{1}
  \rbt^{\frac{1-(-1)^{n_{2i}}}{2}}\lbt\gamma^{0}\rbt^{\frac{1-(-1)^{n_{2i+1}}}{2}}.
\end{align}
Let us remark that the signs in front of $P^{\text{a}}_{\gamma}$ and $\bar{P}^{\text{a}}_{\gamma}$ are the same.
Let $n_{0}^{\text{a}}$ and $n_{1}^{\text{a}}$ the total number of 
$\gamma^{0}$ (resp. $\gamma^{1}$) in $P^{\text{a}}_{\gamma}$. This product
$P^{\text{a}}_{\gamma}$ may be
\begin{enumerate}
\item $\gamma^{0}\gamma^{1}\dotsm\gamma^{0}\gamma^{1}$,
  $n_{0}^{\text{a}}=n_{1}^{\text{a}}$.
  \begin{align}
    P^{\text{a}}_{\gamma}=&
    \begin{cases}
      (-1)^{p}\mathds{1}&\text{if }n_{0}^{\text{a}}=n_{1}^{\text{a}}=2p\\
      (-1)^{p}\gamma^{0}\gamma^{1}&\text{if }n_{0}^{\text{a}}=n_{1}^{\text{a}}=2p+1
    \end{cases}\label{eq:cycle1}
  \end{align}
\item $\gamma^{1}\gamma^{0}\dotsm\gamma^{1}\gamma^{0}$,
  $n_{0}^{\text{a}}=n_{1}^{\text{a}}$.
  \begin{align}
    P^{\text{a}}_{\gamma}=&
    \begin{cases}
      (-1)^{p}\mathds{1}&\text{if }n_{0}^{\text{a}}=n_{1}^{\text{a}}=2p\\
      (-1)^{p}\gamma^{1}\gamma^{0}&\text{if }n_{0}^{\text{a}}=n_{1}^{\text{a}}=2p+1
    \end{cases}\label{eq:cycle2}
  \end{align}
\item $\gamma^{0}\gamma^{1}\dotsm\gamma^{0}\gamma^{1}\gamma^{0}$,
  $n_{0}^{\text{a}}=n_{1}^{\text{a}}+1$.
  \begin{align}
    P^{\text{a}}_{\gamma}=&
    \begin{cases}
      (-1)^{p}\gamma^{0}&\text{if }n_{1}^{\text{a}}=2p\\
      (-1)^{p}\gamma^{1}&\text{if }n_{1}^{\text{a}}=2p+1
    \end{cases}\label{eq:chaine1}
  \end{align}
\item $\gamma^{1}\gamma^{0}\dotsm\gamma^{1}\gamma^{0}\gamma^{1}$,
  $n_{1}^{\text{a}}=n_{0}^{\text{a}}+1$.
  \begin{align}
    P^{\text{a}}_{\gamma}=&
    \begin{cases}
      (-1)^{p}\gamma^{1}&\text{if }n_{0}^{\text{a}}=2p\\
      (-1)^{p}\gamma^{0}&\text{if }n_{0}^{\text{a}}=2p+1
    \end{cases}\label{eq:chaine2}
  \end{align}
\end{enumerate}
Let us apply those results to the chains and cycles of a graph. First of all, remark that the numbers of 
$\gamma^{0}$ and $\gamma^{1}$ in the alternating producthave the same parity as the total numbers in $P_{\gamma}$.
Each cycle contains an even number of $\gamma^{0}$ and $\gamma^{1}$ and then corresponds to situations of the type \eqref{eq:cycle1}
or \eqref{eq:cycle2}. These are exactly symmetric under the exchange $\gamma^{0}\leftrightarrow\gamma^{1}$. When the two chains of a four-point graph contain an odd number of $\gamma^{0}$ and an even number of $\gamma^{1}$, we are faced to situations $3$ or $4$. They are symmetric under the exchange $\gamma^{0}\leftrightarrow\gamma^{1}$. The relative sign between the products $P^{\text{a}}_{\gamma}$ and $\bar{P}^{\text{a}}_{\gamma}$ is $+$ and (especially) only depends on the parities of the total numbers of $\gamma^{0}$ and
$\gamma^{1}$. This sign doesn't depend on the configuration of the products of matrices i.e. it doesn't depend on the $n_{j}$ in \eqref{eq:pgamma}.\\ 

Then the counterterm $\psi\Gamma^{C}\psib\psi\Gamma^{C}\psib$ may only be of the form $\psi\mathds{1}\psib\psi\mathds{1}\psib$,
$\psi\gamma^{0}\gamma^{1}\psib\psi\gamma^{0}\gamma^{1}\psib$ or
$\psi\gamma^{\mu}\psib\psi\gamma_{\mu}\psib$. The result is the same for the two others $\psib\Gamma^{C}\psi\psib\Gamma^{C}\psi$ and
$\psib_{a}\psi_{c}\psib_{b}\psi_{d}\Gamma^{C}_{ab}\Gamma^{C}_{cd}$.
The sum of the last two interactions in \eqref{eq:or2-contreterme} is a linear combination of the initial interactions. We would check it in the same way for $\psib_{a}\psi_{c}\psib_{b}\psi_{d}\Gamma^{C}_{ab}\Gamma^{C}_{cd}$. This proves that $\tau A$ is of the form of the initial vertices.\\

As expected for the four-point function, $\tau A$ is logarithmically divergent. To check it, it is sufficient to redo the procedure used in section \ref{sec:masselottes} with the change of variables (\ref{eq:chgtvarVl}) and \eqref{eq:chgtvarWl} but without $x_{1}$ (the external variables are decoupled form the internal ones in the counerterm). The remainder $(1-\tau)A$ is composed of four different terms. Each improves the power counting and makes $(1-\tau)A$ convergent as $i-e\to\infty$ :
\medskip
\begin{itemize}
\item $\kU\cdot\nabla\delta(\Delta+s\kU)$. Integrating by parts over an external variable, the $\nabla$ acts on an external field and gives at most $M^{e}$. $\kU$ gives at least $M^{-i}$.
\item $XQ_{X\!U}U$, $XQ_{X\!P}P$. $X$ brings $M^{e}$ and $U$ (resp. $P$) $M^{-i}$.
\item $UQ_{U}U$, $PQ_{P}P$ give at least $M^{-2i}$,
\item the expansion \eqref{eq:taylor-propa4} of the propagators gives $M^{-i}$.
\end{itemize}
\medskip
As a conclusion, these termes improves the power counting by $M^{-(i-e)}$ which makes $(1-\tau)A$ convergent and irrelevant for renormalization.

\subsubsection{$B=2$, critical}

The power counting proved in \eqref{eq:compt-bound} let us think that the critical connected components are logarithmically divergent. Exact computations on simple graphs and the behaviour of the theory in the matrix basis confirm this fact. But the divergent part of these graphs are not of the form of the initial Lagrangian and particularly not of a Moyal type. Despite such a divergence, we won't renormalize those graphs. In fact, we will prove in section \ref{critic-comp-2pts} that the renormalization of the corresponding two-point function is sufficient to make the complete graph convergent, including the critical sub-divergence. Let $i$ the scale of the critical component and $j<i$ the scale of the corresponding two-point function. The remainder terms in the renormalization of this two-point function  will give $M^{-(i-e)}$ $(<M^{-(j-e)})$.

\subsection{The two-point function}
\label{subsec:2pt-fct}

\subsubsection{The regular case}
\label{sec:regular-case}

Let a two-point planar subgraph needing renormalization. There exists $(i,k)\in\N^{2}$ such that $N(G^{i}_{k})=2,g(G^{i}_{k})=0$. The two external points of the amputed graph are written $x,y$. The amplitude associated to the connected component $G^{i}_{k}$ is
\begin{align}
  A^{\mu}_{G^{i}_{k}}(x,y)=&\int dxdy\,\psib_{e}(x)\psi_{e}(y)\delta_{G^{i}_{k}} e^{\imath\varphi'_{\Omega}}
  \prod_{l\in\cT^{i}_{k}}du_{l}dv_{l}dp_{l}\,\bar{C}^{i_{l}}_{l}(u_{l})\prod_{\ell\in\cL^{i}_{k}}du_{\ell}dw_{\ell}\,
  \bar{C}^{i_{\ell}}_{\ell}(u_{\ell})\notag
\end{align}
Let us proceed to a second order Taylor expansion. First of all, we expand $\delta_{G^{i}_{k}}$ as
\begin{align}
  \delta_{G^{i}_{k}}\big(x-y+s\kU\big)\Big|_{s=1}=&\delta(x-y)+\kU\cdot\nabla\delta(x-y)+\int_{0}^{1}
  ds\,(1-s)(\kU\cdot\nabla)^{2}\delta(\Delta+s\kU)
\end{align}
where we used the same notations as in the preceding section. The oscillation between $x$ and $y$ is $\exp\imath x\wed y$. Thanks to the delta function, we absorb this oscillation into a redefined matrix $Q_{X\!U}$. Then we expand the oscillation:
\begin{align}
  &\exp\imath(XQ_{X\!U}U+XQ_{X\!P}P+UQ_{U}U+PQ_{P}P)=1+\imath(XQ_{X\!U}U+XQ_{X\!P}P)\\
  &-\int_{0}^{1}ds\,\Big((1-s)(XQ_{X\!U}U+XQ_{X\!P}P)^{2}-\imath(UQ_{U}U+PQ_{P}P)\Big)\notag\\
  &\hspace{1cm}\times e^{\imath s(XQ_{X\!U}U+XQ_{X\!P}P+UQ_{U}U+PQ_{P}P)}.\notag
\end{align} 
We also expand the internal propagators. For all line $l\in G^{i}_{k}$,
\begin{align}
  \bar{C}_{l}(u_{l},s=1)=&\frac{\Omega}{\theta\pi}\int_{0}^{\infty}\frac{dt_{l}\,e^{-t_{l}m^{2}}}{\sinh(2\Ot t_{l})}\,  
  e^{-\frac{\Ot}{2}\coth(2\Ot t_{l})u_{l}^{2}}\big(\imath\Ot\coth(2\Ot
  t_{l})\epsilon(l)\veps(l)\us_{l}\label{eq:taylor-propa2}\\
&\qquad+s\Omega\epsilon(l)\veps(l)\uts_{l}+m\big)\big(\cosh(2\Ot t_{l})\mathds{1}_{2}-s{\textstyle\frac{\theta}{2}}\imath\sinh(2\Ot
  t_{l})\gamma\Theta^{-1}\gamma\big)\Big|_{s=1}\notag\\
  =&\frac{\Omega}{\theta\pi}\int_{0}^{\infty}\frac{dt_{l}\,e^{-t_{l}m^{2}}}{\tanh(2\Ot t_{l})}\,  
  e^{-\frac{\Ot}{2}\coth(2\Ot t_{l})u_{l}^{2}}\lbt\imath\Ot\coth(2\Ot
  t_{l})\epsilon(l)\veps(l)\us_{l}+m\rbt\notag\\
  &+\frac{\Omega}{\theta\pi}\int_{0}^{1}ds\,\int_{0}^{\infty}\frac{dt_{l}\,
    e^{-t_{l}m^{2}}}{\sinh(2\Ot t_{l})}\,e^{-\frac{\Ot}{2}\coth(2\Ot
    t_{l})u_{l}^{2}}\notag\\
  &\qquad\times\Big\{\Omega\epsilon(l)\veps(l)\uts_{l}\big(\cosh(2\Ot
  t_{l})\mathds{1}_{2}-s\imath{\textstyle\frac{\theta}{2}}\sinh(2\Ot
  t_{l})\gamma\Theta^{-1}\gamma\big)\notag\\ 
  &\qquad-\imath{\textstyle\frac{\theta}{2}}\sinh(2\Ot t_{l})\big(\imath\Ot\coth(2\Ot
  t_{l})\epsilon(l)\veps(l)\us_{l}+s\Omega\epsilon(l)\veps(l)\uts_{l}+m\big)\gamma\Theta^{-1}\gamma\Big\}.\notag
\end{align}
The conscientious reader would have noticed that the expansion
\eqref{eq:taylor-propa2} is different from the one we used for the four-point function \eqref{eq:taylor-propa4}. Here we allow the mass term to be part of the zeroth order term. The reason is that the number of internal lines in a two-point function is odd (it is $2n-1$). For the mass counterterm, if all the propagators would have contributed by a $u$ term, the counterterm woud have vanished. In fact, the power counting is reached when one propagator uses its mass term and all the others the term $\us$. This implies that the mass divergence is only logarithmic. For the wave function and $\Omega\xts$ counterterm, each propagator contributes with its dominant term $\us$. The counterterm $\tau A$ associated to the connected component $G^{i}_{k}$ corresponds to the zeroth and first order terms of the three preceding expansions:
\begin{align}
  \tau A^{\mu}_{G^{i}_{k}}=\phantom{\int}&\negthickspace\!\!\tau A_{m}+\tau A_{\ps}+\tau A_{\xts},\\
  \tau A_{m}=\int&dxdy\,\psib_{e}(x)\psi_{e}(y)\delta(x-y)
  \int\prod_{l\in\cT^{i}_{k}}du_{l}dv_{l}dp_{l}\,\bar{C}^{i_{l}}_{l}(u_{l},s=0)\label{eq:massterm}\\
  &\prod_{\ell\in\cL^{i}_{k}}du_{\ell}dw_{\ell}\,
  \bar{C}^{i_{\ell}}_{\ell}(u_{\ell},s=0)e^{\imath\varphi'_{\Omega}(s=0)},\notag\\
  \tau A_{\ps}=\int&dxdy\,\psib_{e}(x)\psi_{e}(y)\kU\cdot\nabla\delta(x-y)
  \prod_{l\in\cT^{i}_{k}}du_{l}dv_{l}dp_{l}\,\bar{C}^{i_{l}}_{l}(u_{l},s=0)\label{eq:waveterm}\\
  &\prod_{\ell\in\cL^{i}_{k}}du_{\ell}dw_{\ell}\,
  \bar{C}^{i_{\ell}}_{\ell}(u_{\ell},s=0)e^{\imath\varphi'_{\Omega}(s=0)},\notag\\
  \tau A_{\xts}=\imath\int&dxdy\,\psib_{e}(x)\psi_{e}(y)\delta(x-y)(XQ_{X\!U}U+XQ_{X\!P}P)\label{eq:Omegaterm}\\
  &\prod_{l\in\cT^{i}_{k}}du_{l}dv_{l}dp_{l}\,\bar{C}^{i_{l}}_{l}(u_{l},s=0)
  \prod_{\ell\in\cL^{i}_{k}}du_{\ell}dw_{\ell}\,
  \bar{C}^{i_{\ell}}_{\ell}(u_{\ell},s=0)e^{\imath\varphi'_{\Omega}(s=0)}.\notag
\end{align}
The counterterm $\tau A_{m}$ contributes to the mass renormalization. Its divergence is logarithmic for the parity reasons given above. $\tau A_{\ps}$ is the wave function counterterm.
\begin{align}
  \tau A_{\ps}=-\int&dx\,\psib_{e}(x)\nabla^{\mu}\psi_{e}(x)\kU^{\mu}
  \prod_{l\in\cT^{i}_{k}}du_{l}dv_{l}dp_{l}\,\bar{C}^{i_{l}}_{l}(u_{l},s=0)\\
  &\prod_{\ell\in\cL^{i}_{k}}du_{\ell}dw_{\ell}\,
  \bar{C}^{i_{\ell}}_{\ell}(u_{\ell},s=0)e^{\imath\varphi'_{\Omega}(s=0)}\notag
\end{align}
As for the four-point function, this term contains the polynomial \eqref{eq:polyn} here of odd degree. The gamma matrices in each monomial are distributed among cycles and a chain (see definition \ref{def:cycle-chaine}). The numbers of $u^0\gamma^{0}$ and $u^1\gamma^{1}$ in each cycle are even so that the number of gamma matrices in the chain linking the external points is odd. The term $\psib_{e}\kU^{0}\partial_{0}\psi_{e}$ is different from zero if the number of $u^{0}\gamma^{0}$ in the chain is odd. Then the number of $\gamma^{1}$ is even. The corresponding counterterm is of the form $\psib_{e}\gamma^{0}\partial_{0}\psi_{e}$. We associate it the term $\psib_{e}\kU^{1}\partial_{1}\psi_{e}$ where we chose the inverse monomial in $P$ ($\forall l\in G,\,
\gamma^{0}u_{l}^{0}\leftrightarrow\gamma^{1}u_{l}^{1}$). Thanks to a rotation of the coordinates, we show that the complete counterterm looks like $\psib_{e}\slashed{\nabla}\psi_{e}$. It is logarithmically divergent.\\

The counterterm $\tau A_{\xts}$, also logarithmically divergent, contributes to the renormalization of the ``magnetic field'' $\Omega\xts$. The terms entering such a contribution look like $\int\psib_{e}\psi_{e}(x^{0}u^{1}-x^{1}u^{0})\dotsm$. Once more we can associate two opposite monomials and perform a rotation to prove that the counterterm is of the form $\psib_{e}\xts\psi_{e}$. Remark that the terms $\int\psib_{e}\psi_{e}(x^{0}p_{0}+x^{1}p_{1})\dotsm$ vanish by parity over $p$ (beware that here $p_{\mu}$ is the ``momentum'' associated to a tree line and not a derivative). It is easy to check from \eqref{eq:waveterm} and \eqref{eq:Omegaterm} that the counterterms $\tau A_{\ps}$ and $\tau A_{\xts}$ are skew-Hermitian. They are of the form $\psib\ps\psi$ and $\psib\xts\psi$.\\

The remainder terms, gathered in $(1-\tau)A$, are convergent:
\begin{itemize}
\item $(\kU\cdot\nabla)^{2}\delta$ gives $M^{-2i}$ thanks to $\kU^{2}$ and
  $M^{2e}$ by integration by parts over an external point,
\item $(XQ_{X\!U}U+XQ_{X\!P}P)^{2}$ brings $M^{-2(i-e)}$,
\item $UQ_{U}U+PQ_{P}P$ give at least $M^{-2i}$,
\item The propagators expansion gives at least $M^{-i}$.
\end{itemize}
Note that until now the $\psib\gamma^{0}\gamma^{1}\psi$ counterterm was not useful. Moreover if we set $m=0$ (the bare mass) it remains so under radiative correction ($\tau A_{m}\equiv 0$) for parity reasons over the $u$'s.

\subsubsection{Critical components}
\label{critic-comp-2pts}

Let us consider an orientable two-point graph at scale $j$ with a critical subgraph at scale $i>j$ (see definition in section \ref{sec:multiscaleGN}). This two-point component is then made of a four-point subgraph at a scale $i$ with $g=0, B=2$ and of a single (loop) line of scale $j$. We renormalize the two-point amplitude as was done in the previous paragraph. We now want to show that the remainder terms are of order $M^{-2(i-e)}$ (and not $M^{-2(j-e)}$) which implies the convergence of the complete remainder amplitude even its four-point sub-divergence.\\

We proceed as is explained in section \ref{sec:multiscaleGN}. Down to scale $i$, we get all the necessary masslets for the $v$'s and $w$'s and the corresponding functions for the $p$'s. Then we have an oscillation $\cW_{\ell}\wed u^{j}$ where $i_{\ell}=i$ and $u^{j}$ is the $u$ variable of the unique loop line of scale $j$. We use it to get a decreasing function $\ks$ implementing $\labs u^{j}\rabs\les M^{-i}$. It remains to obtain the masslet for the variable $w^{j}$. Its associated $u^{j}$ variable being now of order $M^{-i}$ there is no mean to get a masslet of scale $M^j$. We can only achieve $M^{i}$. The gain we had with the $u^{j}$ variable is lost by its corresponding masslet and we note once again that the critical components are divergent. But now all the $u$ variables in the graph are bounded by $M^{-i}$ which implies that the remainder terms, except the propagator expansions, bring $M^{-2(i-e)}=M^{-2(i-j)}M^{-2(j-e)}$. All the propagator expansions except the one concerning the lowest propagator (of scale $j$) give at least $M^{-i}$. There is one term in the expansion of the lowest propagator ($\imath m{\textstyle\frac{\theta}{2}}\sinh(2\Ot t_{\ell})\gamma\Theta^{-1}\gamma$) which only brings $M^{-2j}$. This is not sufficient to renormalize the four-point sub-divergence. The solution consists in putting that term in the counterterm. Only for this lowest propagator, we use a different propagator expansion:
\begin{align}
  \bar{C}_{l}(u_{l},s=1)=&\frac{\Omega}{\theta\pi}\int_{0}^{\infty}\frac{dt_{l}\,e^{-t_{l}m^{2}}}{\sinh(2\Ot t_{l})}\,  
  e^{-\frac{\Ot}{2}\coth(2\Ot t_{l})u_{l}^{2}}\big(\imath\Ot\coth(2\Ot
  t_{l})\epsilon(l)\veps(l)\us_{l}\label{eq:taylor-propa-crit}\\
&\qquad+s\Omega\epsilon(l)\veps(l)\uts_{l}+m\big)\big(\cosh(2\Ot t_{l})\mathds{1}_{2}-{\textstyle\frac{\theta}{2}}\imath\sinh(2\Ot
  t_{l})\gamma\Theta^{-1}\gamma\big)\Big|_{s=1}\notag\\
  =&\frac{\Omega}{\theta\pi}\int_{0}^{\infty}\frac{dt_{l}\,e^{-t_{l}m^{2}}}{\tanh(2\Ot t_{l})}\,  
  e^{-\frac{\Ot}{2}\coth(2\Ot t_{l})u_{l}^{2}}\lbt\imath\Ot\coth(2\Ot
  t_{l})\epsilon(l)\veps(l)\us_{l}+m\rbt\notag\\
  &\qquad\times\big(\cosh(2\Ot t_{l})\mathds{1}_{2}-{\textstyle\frac{\theta}{2}}\imath\sinh(2\Ot
  t_{l})\gamma\Theta^{-1}\gamma\big)\notag\\
  &+\frac{\Omega}{\theta\pi}\int_{0}^{1}ds\,\int_{0}^{\infty}\frac{dt_{l}\,
    e^{-t_{l}m^{2}}}{\sinh(2\Ot t_{l})}\,e^{-\frac{\Ot}{2}\coth(2\Ot
    t_{l})u_{l}^{2}}\notag\\
  &\qquad\times\Omega\epsilon(l)\veps(l)\uts_{l}\big(\cosh(2\Ot
  t_{l})\mathds{1}_{2}-\imath{\textstyle\frac{\theta}{2}}\sinh(2\Ot
  t_{l})\gamma\Theta^{-1}\gamma\big).\notag 
\end{align}
This makes convergent the four-point subgraph and the two-point one. The price to pay is a counterterm of the form $\imath\delta m\,\theta\gamma\Theta^{-1}\gamma$. The proof of this last statement is given in appendix \ref{sec:modif-count-two}. Remark finally that if we set $m=0$, $\tau A_{m}\equiv 0$ and no $\psib\gamma^{0}\gamma^{1}\psi$ appear.

\section{Conclusion}
\label{sec:conclusion}

We proved that the \nc{} Gross-Neveu model, defined by the action (\ref{eq:actfunct}) with only orientable interactions, is renormalizable to all orders. We have first computed a bound on the amputed amplitude of any graph, integrated over test functions (see lemma \ref{lem:compt-puiss}). This power counting is the one of a renormalizable theory. This bound can be obtained at $\Omega =0$. Then we showed that all the necessary counterterms are of the form of the initial Lagrangian. This means that the \nc{} Gross-Neveu model with orientable interactions is renormalizable even without the vulcanization procedure. But without general argument in favour of orientable interactions, we have to consider also non-orientable ones and then to vulcanize the Lagrangian.

The orientable Gross-Neveu model is free of (non-renormalizable) UV/IR mixing \cite{MiRaSe,CheRoi}. Nevertheless it exhibits some remaining one. It concerns some graphs of the four-point function. These ones have $g=0$ and $B=2$ (see lemma \ref{lem:compt-puiss}). This mixing is fortunately renormalizable in the following sense: the divergent part of the critical four-point graphs is not ``local'' but the renormalization of the corresponding two-point function makes those four-point subgraphs finite. Of course this was not the case for the usual UV/IR mixing which prevented renormalization of \nc{} field theory before \cite{GrWu04-3}. Finally note that the bounds in lemma \ref{lem:compt-puiss} may have equally been proved for the full model (with $V=V_{\text{o}}+V_{\text{no}}$) but restricted to orientable graphs\footnote{Orientable interactions only lead to orientable graphs but orientable graphs are not only made of orientable interactions. Actually non-orientable interactions produce not only all the non-orientable graphs but also orientable ones.}. This suggests that the full theory could be renormalizable if restricted to orientable graphs. Of course the ``locality'' of the counterterms should be checked.

\appendix

\section{Topology of Feynman graphs}
\label{sec:topol-feynm-graphs}

Let a graph G with $n$ vertices and $I$ internal lines. Interactions of quantum field theories on moyal spaces are only cyclically invariant (see (\ref{eq:interaction-phi4})). A good way to keep track of such a reduced invariance is to draw Feynman graphs as ribbon graphs. Moreover there exists a basis for the Schwartz class functions where the Moyal product becomes an ordinary matrix product \cite{GrWu03-2,Gracia-Bondia1987kw}. This further justifies the ribbon representation.\\

\begin{figure}[htbp]
  \centering 
  \subfloat[$x$-space representation]{{\label{fig:x-rep}}\includegraphics[scale=.8]{figures.48}}\qquad
  \subfloat[Ribbon representation]{\label{fig:ribbon}\includegraphics[scale=1]{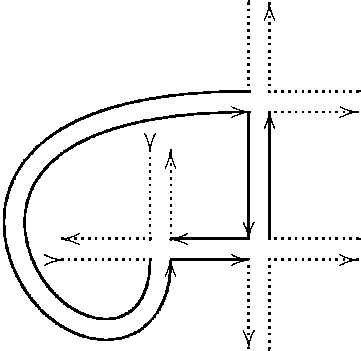}}
  \caption{A graph with two broken faces}
  \label{fig:broken-ex}
\end{figure}

Let us consider the example of figure \ref{fig:broken-ex}.
Propagators in a ribbon graph are made of double lines. Let us call $L$ the number of loops (made of single lines) of a ribbon graph. The graph of figure \ref{fig:ribbon} has $n=3, I=3, L=2$. Each ribbon graph can be drawn on a manifold of genus $g$. The genus is computed from the Euler characteristic $\chi=L-I+n$. For example, the graph of figure \ref{fig:ribbon} may be drawn on a manifold of genus $0$. Note that some of the $L$ loops of a graph may be ``broken'' by external legs. In our example, both loops are broken.

\section{Integration by parts}
\label{sec:integration-parts}
We reproduce here the details of the computation showing that the procedure formed by the change of variables
\eqref{eq:chgtvarVl} and the integration by parts \eqref{eq:int-part} allows to
get a decreasing function of the desired scale.
\begin{align}
  A_{G,l}=\int&
  da_{l}dt_{l}\,\coth(2\Ot
  t_{l})\xi(a_{l}\coth^{1/2}(2\Ot t_{l}))\,
  e^{-\frac{\Ot}{2}\coth(2\Ot
    t_{l})(u_{l}-\veps(l)a_{l})^{2}}f_{1}(x_{1}+\eta(1)\veps(l)a_{l})\notag\\
  &\hspace{-.3cm}\lb\imath\Ot\coth(2\Ot
  t_{l})(\epsilon\veps)(l)(\us_{l}-\veps(l)\slashed{a}_{l})+\Omega(\epsilon\veps)(l)
  (\uts_{l}-\veps(l)\slashed{\tilde{a}}_{l})-m\rb e^{\imath a_{l}\wed(U_{l}+A_{l}+X_{l})}\nonumber\\
  &\prod_{\mu=0}^{1}\lbt\frac{\coth^{1/2}(2\Ot t_{l})+\frac{\partial}{\partial
      a_{l}^{\mu}}}{\coth^{1/2}(2\Ot
    t_{l})+\imath\cVt_{l,\mu}}\rbt^{\!\!\!2} e^{\imath a_{l}\wed\cV_{l}}\tag*{\eqref{eq:int-part}}
  \intertext{Let us write $\kc_{l}=\coth(2\Ot t_{l})$.}
  A_{G,l}=\int&da_{l}\,\kc_{l}\, e^{\imath
    a_{l}\wed\cV_{l}}\prod_{\mu=0}^{1}\lbt\frac{1}{\sqrt{\kc_{l}}+\imath\cVt_{l,\mu}}\rbt^{2}\lbt\sqrt{\kc_{l}}
  -\frac{\partial}{\partial
    a_{l}^{\mu}}\rbt^{2}\xi(a_{l}\sqrt{\kc_{l}})f_{1}(x_{1}+\eta(1)\veps(l)a_{l})\notag\\
  &\hspace{-1.2cm}\lb\imath\Ot\kc_{l}(\epsilon\veps)(l)(\us_{l}-\veps(l)\slashed{a}_{l})+\Omega(\epsilon\veps)(l)
  (\uts_{l}-\veps(l)\slashed{\tilde{a}}_{l})-m\rb e^{-\frac{\Ot}{2}
    \kc_{l}(u_{l}-\veps(l)a_{l})^{2}+\imath a_{l}\wed(U_{l}+A_{l}+X_{l})}.\notag
\end{align}
We define the following notations:
\begin{align}
  \lb l\rb=&\imath\Ot\kc_{l}(\epsilon\veps)(l)(\us_{l}-\veps(l)\slashed{a}_{l})
  +\Omega(\epsilon\veps)(l)(\uts_{l}-\veps(l)\slashed{\tilde{a}}_{l})-m,\\
  \lb l\rb'\!=&-\epsilon(l)\lbt\imath\Ot\kc_{l}\gamma^{\mu}+\Ot(-1)^{\mu+1}\gamma^{\mu+1}\rbt.
\end{align}
Let us compute the first derivative:
\begin{align}
  &\frac{\partial}{\partial
    a_{l}^{\mu}}\,e^{-\frac{\Ot}{2}
    \kc_{l}(u_{l}-\veps(l)a_{l})^{2}+\imath a_{l}\wed(U_{l}+A_{l}+X_{l})}\xi(a_{l}\sqrt{\kc_{l}})f_{1}(x_{1}+\eta(1)\veps(l)a_{l})\lb l\rb\\
  =&e^{-\frac{\Ot}{2}\kc_{l}(u_{l}-\veps(l)a_{l})^{2}+\imath a_{l}\wed(U_{l}+A_{l}+X_{l})} \Big\{\{l\}\big[\Ot\kc_{l}\veps(l)(u_{l}-\veps(l)a_{l})^{\mu}\xi f_{1}
  +\imath(\Ut_{l}+\At_{l}+\Xt_{l})_{\mu}\xi f_{1}\notag\\
  &+\sqrt{\kc_{l}}\xi'f_{1}+\eta(1)\veps(l)\xi f'_{1}\big]+\lb l'\rb\xi f_{1}\Big\}.\notag
\end{align}
Then the second one:
\begin{align}
    &\frac{\partial^{2}}{\partial
    (a_{l}^{\mu})^{2}}\,e^{-\frac{\Ot}{2}
    \kc_{l}(u_{l}-\veps(l)a_{l})^{2}+\imath a_{l}\wed(U_{l}+A_{l}+X_{l})}\xi(a_{l}\sqrt{\kc_{l}})f_{1}(x_{1}+\eta(1)\veps(l)a_{l})\lb l\rb\\
  =&e^{-\frac{\Ot}{2}\kc_{l}(u_{l}-\veps(l)a_{l})^{2}+\imath a_{l}\wed(U_{l}+A_{l}+X_{l})} \Big\{\{l\}\big[\big(\Ot\kc_{l}\veps(l)(u_{l}-\veps(l)a_{l})^{\mu}\xi f_{1}
  +\imath(\Ut_{l}+\At_{l}+\Xt_{l})_{\mu}\xi f_{1}\notag\\
  &+\sqrt{\kc_{l}}\xi'f_{1}+\eta(1)\veps(l)\xi f'_{1}\big)\times\big(\Ot\kc_{l}\veps(l)(u_{l}-\veps(l)a_{l})^{\mu}\xi f_{1}
  +\imath(\Ut_{l}+\At_{l}+\Xt_{l})_{\mu}\xi f_{1}\big)\notag\\
  &-\Ot\kc_{l}\veps(l)\xi f_{1}+\Ot\kc_{l}^{3/2}(u_{l}-\veps(l)a_{l})^{\mu}\xi'f_{1}+\Ot\eta(1)\veps(l)\kc_{l}(u_{l}-\veps(l)a_{l})^{\mu}\xi f'_{1}\notag\\
  &+\imath\sqrt{\kc_{l}}(\Ut_{l}+\At_{l}+\Xt_{l})_{\mu}\xi' f_{1}+\imath\eta(1)\veps(l)(\Ut_{l}+\At_{l}+\Xt_{l})_{\mu}\xi f'_{1}\notag\\
  &+\kc_{l}\xi''f_{1}+2\sqrt(\kc_{l})\eta(1)\veps(l)\xi'f'_{1}+\xi f''_{1}\big]\notag\\
  &+2\{l'\}\big[\Ot\kc_{l}\veps(l)(u_{l}-\veps(l)a_{l})^{\mu}\xi f_{1}
  +\imath(\Ut_{l}+\At_{l}+\Xt_{l})_{\mu}\xi f_{1}+\sqrt(\kc_{l})\xi'f_{1}+\eta(1)\veps(l)\xi f'_{1}\big]\Big\}.\notag
\end{align}
The terms we get are of order $\cO(\kc_{l}^{3/2})$. This gives \eqref{eq:intpart-result}.

\section{The vacuum graphs}
\label{sec:vacuum-graph}

In this appendix, we compute the power counting of the vacuum graphs of the orientable Gross-Neveu model. Let us first remind that the translation invariance of the usual commutative field theories makes them infinite even with both ultraviolet and infrared cut-offs (we mean in a given slice). On the contrary, the vacuum graphs of the (\nc{}) $\Phi^{4}$ theory are finite in a slice but the sum over their scale attribution diverges as $M^{8i}$.\\

The quartic Moyal-type interaction is translation invariant. It can indeed be written as
\begin{align}
  &\delta(x_{1}-x_{2}+x_{3}-x_{4})\exp\imath\sum_{i<j=1}^{4}(-1)^{i+j+1}x_{i}\wed x_{j}\\
  =&\delta(x_{1}-x_{2}+x_{3}-x_{4})\exp\imath\lbt
  x_{1}\wed(x_{2}-x_{3})+x_{2}\wed x_{3}\rbt\nonumber\\
  =&\delta(x_{1}-x_{2}+x_{3}-x_{4})\exp\imath(x_{1}-x_{2})\wed(x_{2}-x_{3})\nonumber
\end{align}
Such a regularisation is then solely due to the breakdown of translation invariance by the harmonic potential term $\xt^{2}$ in the $\Phi^4$ propagator. The Gross-Neveu propagator, whereas breaking tranlation invariance, allows to get translation invariant amplitudes for the vacuum graphs. We verify such an invariance by performing the change of variables $\forall i,\, x_{i}\to x_{i}+a$ and by checking that the result is independant $a$.
\begin{align}
  A_{G}=&\lambda^{n}\int\prod_{l\in G}du_{l}dv_{l}\,C_{l}(u_{l},v_{l})\
  e^{\imath\varphi}\label{eq:AGN}\\
  =&\lambda^{n}\int\prod_{l\in G}du_{l}dv_{l}\,C_{l}(u_{l},v_{l}+2a)\
  e^{\imath\varphi}\nonumber
\end{align}
In equation (\ref{eq:AGN}), we wrote $v_{l}$ for all lines to simplify notations. We have already noticed that the vertex oscillations are translation invariant. That's why under the change of variables, $\varphi$ remains unchanged. Let us consider a $\psib\psi\psib\psi$ type interaction. In that case, the popagator oscillations are always $\exp-\frac{\imath\Omega}{2}u_{l}\wed v_{l}$. Then the change of variables $v_{l}\to v_{l}+2a_{l}$ implies the following $a$ dependance for any amplitude
\begin{equation}
  \label{eq:adepGN}
  \exp\imath\Omega a\wed\sum_{l\in G}u_{l}=1
\end{equation}
wich is $1$ because the sum of all the $u$ variables vanishes for the vacuum graphs thanks to the root delta function (\ref{eq:deltaroot}) (remind that we only consider orientable interactions). This proves that the vacuum graphs of the orientable Gross-Neveu model are infinite.\\
\begin{figure}[htbp]
  \centering 
  \includegraphics[scale=1]{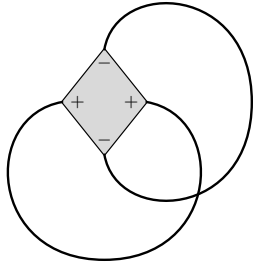}
  \caption[Vacuum graph]{Example of non-orientable vacuum graph}
  \label{fig:no-vacuum-ex}
\end{figure}
For non-orientable interaction, this is not the case as the reader may verify on the example of figure \ref{fig:no-vacuum-ex}.

%%%%%%%%%
\section{(Un)Modified counterterms of the two-point\\function}
\label{sec:modif-count-two}

Let us consider a two-point connected component $G^{j}_{k'}$ with a critical sub-divergent component $G^{i}_{k}$. We prove that, if we put the $\gamma\Theta^{-1}\gamma$ term of the lowest propagator $\ell_{0}$ in $G^{j}_{k'}$ into the counterterm, the divergent part of this two-point function remains of the form of the initial Lagrangian.

For simplicity we use a lightened notation than until now: $\exp-2\imath\Omega t_{\ell_{0}}\gamma\Theta^{-1}\gamma=\cosh(2\Ot t_{\ell_{0}})\bbbone_{2}-\imath\sinh(2\Ot t_{\ell_{0}})\gamma^{0}\gamma^{1}$. As explained in section \ref{subsec:4pt-fct}, the propagators in a two-point function are distributed among cycles and a chain. For any given graph $G$, let us write $\bC$ for the set of all cycles and $\bCh$ for the set of all chains. We also write $T^{\mu}$ for the number of $u^{\mu}$'s coming from the Taylor expansions\footnote{For example, for the mass term, the Taylor expansion brings no $u$'s then $T^0=T^1=0$. The wave function counterterm brings $u^0\partial_{0}+u^1\partial_{1}$. The first term has $T^0=1$ and $T^1=0$, the second the contrary.}. Each cycle or chain consists in a product of propagators. Let $c\in\bC\,(\bCh)$,
\begin{align}
  P_{c}=&
  \begin{cases}
    \prod_{l\in c}(\imath\Ot\coth(2\Ot t_{l})\us_{l}+m)&\text{if }\ell_{0}\notin c\\
    (\imath\Ot\coth(2\Ot t_{\ell_{0}})\us_{\ell_{0}}+m)e^{-2\imath\Ot t_{\ell_{0}}\gamma^{0}\gamma^{1}}\prod_{l\in c\setminus\{\ell_{0}\}}(\imath\Ot\coth(2\Ot t_{l})\us_{l}+m)&\text{if }\ell_{0}\in c.
  \end{cases}
\end{align}
$P_{c}$ is a sum of different terms: $P_{c}=\sum_{i=1}^{n}P_{c}^{i}$ where $n=2^{|c|}$ if $c\in\bC$ and $n=2^{|c|+1}$ if $c\in\bCh$ ($|c|=\card c$). Let us write $|\gamma^{\mu}|_{c}^{i}$ for the total number of $\gamma^{\mu}$ in a given term $i$ of $c\in\bC\,(\bCh)$. In the same way, we define $|u^{\mu}|_{c}^{i}$. Let $i_{c}\in\lnat 1,n\rnat$ for all $c\in\bC\cup\bCh$. The tracelessness of the gamma matrices and the parity properties of the integrales over the $u$'s implies two constraints:
\begin{align}
  \forall c\in\bC,\,\forall i\in\lnat 1,2^{|c|}\rnat,\,\forall\mu\in\{0,1\},\,|\gamma^{\mu}|^{i}_{c} \text{ is}&\text{ even},\label{eq:cycle-constr}\\
  \forall\mu\in\{0,1\},\,\sum_{c\in\bC\cup\bCh}|u^{\mu}|^{i_{c}}_{c}+T^{\mu}\text{ is}&\text{ even.}\label{eq:chain-constr}
\end{align}
From now on, we fix a $\N$-valued sequence $(i_{c})_{c\in\bC\cup\bCh}$. Remind that in a two-point function, $|\bCh|=1$ and that the total number of internal lines is odd: $\sum_{c\in\bC\cup\bCh}|c|$ is odd. For $\ell_{0}$, we will always choose its $\gamma^{0}\gamma^{1}$ term otherwise the analysis is the same as in section \ref{subsec:2pt-fct}. In the following we call ``mass counterterm'' the expression (\ref{eq:massterm}) with the expansion (\ref{eq:taylor-propa-crit}), ``$\ps$ (or $\xts$) counterterm'' the equation (\ref{eq:waveterm}) (or (\ref{eq:Omegaterm})) once more with the expansion (\ref{eq:taylor-propa-crit}). 
\begin{enumerate}
\item Let $c_{1}\in\bCh$. If $|c_{1}|$ (the number of lines in the chain) is even
  \begin{enumerate}
  \item and $\ell_{0}\in c_{1}$, $\sum_{c\in\bC}|c|$ is odd. Equation (\ref{eq:cycle-constr}) implies $\forall\mu,\,\sum_{c\in\bC}|u^{\mu}|^{i_{c}}_{c}$ even. The total number of lines in the cycles being odd, we chose the mass for at least one line in $\bC$.
    \begin{itemize}
    \item For the mass counterterm, $T^0=T^1=0$. Equation (\ref{eq:chain-constr}) implies $|u^{\mu}|^{i_{c_{1}}}_{c_{1}}$ even. This gives
      $|\gamma^{\mu}|^{i_{c_{1}}}_{c_{1}}$ both odd. The counterterm may only be proportionnal to $\gamma^0\gamma^1$.
    \item For the $\ps$ or $\xts$ counterterm, let $\mu\in\Z_{2},\,T^\mu=1$ and $T^{\mu+1}=0$. $|\gamma^{\mu}|^{i_{c_{1}}}_{c_{1}}$ is even and $|\gamma^{\mu+1}|^{i_{c_{1}}}_{c_{1}}$ is odd. The number of lines in $c_{1}$ being even, at least one line in $c_{1}$ ``chose'' the mass. Then this term is of order $M^{-i}$. Such terms give $\pts$ or $\xs$.
    \end{itemize}
  \item Let $\ell_{0}\notin c_{1}$. Equation (\ref{eq:cycle-constr}) implies $\forall\mu,\,\sum_{c\in\bC}|u^{\mu}|^{i_{c}}_{c}$ odd. We chose the mass term at least once.
    \begin{itemize}
    \item Mass counterterm: $|u^{\mu}|^{i_{c_{1}}}_{c_{1}}$ is odd. This counterterm is proportionnal to $\gamma^{0}\gamma^{1}$.
    \item $\ps$ ($\xts$) counterterm: $|\gamma^{\mu}|^{i_{c_{1}}}_{c_{1}}$ is odd and $|\gamma^{\mu+1}|^{i_{c_{1}}}_{c_{1}}$ is even. This term gives $\ps$ or $\xts$ but is convergent as $M^{-i}$ since $|c_{1}|$ is even and at least one line in $c_{1}$ bears a mass term. 
    \end{itemize}
  \end{enumerate}
\item If $|c_{1}|$ is odd
  \begin{enumerate}
  \item Let $\ell_{0}\in c_{1}$. $\sum_{c\in\bC}|u^{\mu}|_{c}^{i_{c}}$ is even. 
    \begin{itemize}
    \item Mass counterterm: $|\gamma^{\mu}|^{i_{c_{1}}}_{c_{1}}$'s are both odd. This gives $\psib\gamma^{0}\gamma^{1}\psi$.
    \item $\ps$ ($\xts$) counterterm: $|\gamma^{\mu}|^{i_{c_{1}}}_{c_{1}}$ is even and $|\gamma^{\mu+1}|^{i_{c_{1}}}_{c_{1}}$ is odd. This term gives $\pts$ or $\xs$ but is convergent as $M^{-(i-j)}$. The number of lines in $c_{1}$ being odd, either all the lines in $c_{1}$ chose the $u$ term or at least two of them chose the mass term. 
    \end{itemize}
  \item Let $\ell_{0}\notin c_{1}$. $\sum_{c\in\bC}|\gamma^{\mu}|^{i_{c}}_{c}$'s are both odd. Either all the lines in $\bC$ chose the $u$ term (the total number of lines in $\bC$ is even) or at least two of them chose the mass term. The corresponding terms are of order $M^{-(i-j)}$.
    \begin{itemize}
    \item Mass counterterm: $|\gamma^{\mu}|^{i_{c_{1}}}_{c_{1}}$'s are both odd. We get $\psib\gamma^{0}\gamma^{1}\psi$.
    \item $\ps$ ($\xts$) counterterm: $|\gamma^{\mu}|^{i_{c_{1}}}_{c_{1}}$ is even and $|\gamma^{\mu+1}|^{i_{c_{1}}}_{c_{1}}$ is odd. This term gives $\pts$ or $\xs$.
    \end{itemize}
  \end{enumerate}
\end{enumerate}
As a conclusion, the mass term only brings $\psib\gamma^{0}\gamma^{1}\psi$. The $\ps$ and $\xts$ counterterms may give $\psib\pts\psi$ and $\psib\xs\psi$, not present in the initial Lagrangian, but these terms are convergent and may be let in the remainder term. A way to define the new counterterms is
\begin{align}
  \tau'A_{m}=&\frac 12\Tr(\tau A_{m}),\\
  \tau'A_{\delta m}=&-\frac 12\gamma^{0}\gamma^{1}\Tr(\gamma^{0}\gamma^{1}\tau A_{m}),\\
  \tau'A_{\ps}=&-\frac{\ps}{2p^{2}}\Tr(\ps\tau A_{\ps}),\\
  \tau'A_{\xts}=&-\frac{\xts}{2\xt^{2}}\Tr(\xts\tau A_{\xts}).
\end{align}
Remark that if $m=0$, $\tau A_{m}\equiv 0$. This means that if the bare mass is zero, it remains zero after radiative corrections and no $\psib\gamma^{0}\gamma^{1}\psi$ appear.
%%%%%%%%%%%%%%%%%%%%%%%%%%%
\bibliographystyle{mybibstyle}
\bibliography{biblio-articles,biblio-books}

%%% Local Variables: 
%%% mode: latex
%%%% TeX-command-default : "PDFLaTeX"
%%% case-fold-search : nil
%%% TeX-master: t
%%% End: 

\end{document}